\documentclass[prd,aps,showpacs,epsf,floats,onecolumn]{revtex4}%
\usepackage{amssymb}
\usepackage{amsfonts}
\usepackage{amsmath}
\usepackage{graphicx}%
\providecommand{\U}[1]{\protect\rule{.1in}{.1in}}
\begin{document}
\title{\textbf{From Entanglement to Universality: A Multiparticle Spacetime Algebra
Approach to Quantum Computational Gates Revisited }}
\author{\textbf{Carlo Cafaro}$^{1,2}$, \textbf{Newshaw Bahreyni}$^{3}$,
\textbf{Leonardo Rossetti}$^{4,1}$}
\affiliation{$^{1}$University at Albany-SUNY, Albany, NY 12222, USA}
\affiliation{$^{2}$SUNY Polytechnic Institute, Utica, NY 13502, USA}
\affiliation{$^{3}$Pomona College, Claremont, CA 91711, USA}
\affiliation{$^{4}$University of Camerino, I-62032 Camerino, Italy}

\begin{abstract}
Alternative mathematical explorations in quantum computing can be of great
scientific interest, especially if they come with penetrating physical
insights. In this paper, we present a critical revisitation of our geometric
(Clifford) algebras (GAs) application in quantum computing as originally
presented in [C. Cafaro and S. Mancini, Adv. Appl. Clifford Algebras
\textbf{21}, 493 (2011)]. Our focus is on testing the usefulness of geometric
algebras (GAs) techniques in two applications to quantum computing. First,
making use of the geometric algebra of a relativistic configuration space
(a.k.a., multiparticle spacetime algebra or MSTA), we offer an explicit
algebraic characterization of one- and two-qubit quantum states together with
a MSTA description of one- and two-qubit quantum computational gates. In this
first application, we devote special attention to the concept of entanglement,
focusing on entangled quantum states and two-qubit entangling quantum gates.
Second, exploiting the previously mentioned MSTA\ characterization together
with the GA depiction of the Lie algebras $\mathrm{SO}\left(  3;%
\mathbb{R}
\right)  $ and $\mathrm{SU}\left(  2;%
\mathbb{C}
\right)  $ depending on the rotor group $\mathrm{Spin}^{+}\left(  3,0\right)
$ formalism, we focus our attention to the concept of universality in quantum
computing by reevaluating Boykin's proof on the identification of a suitable
set of universal quantum gates. At the end of our mathematical exploration, we
arrive at two main conclusions. Firstly, the MSTA perspective leads to a
powerful conceptual unification between quantum states and quantum operators.
More specifically, the \emph{complex} qubit space and the \emph{complex} space
of unitary operators acting on them merge in a single multivectorial
\emph{real} space. Secondly, the GA viewpoint on rotations based on the rotor
group $\mathrm{Spin}^{+}\left(  3,0\right)  $ carries both conceptual and
computational upper hands compared to conventional vectorial and matricial methods.

\end{abstract}

\maketitle

\section{Introduction}

A universal mathematical language for physics is the so-called Geometric
(Clifford) algebra (GA) \cite{hestenes, dl}, a language that relies on the
mathematical formalism of Clifford algebra. A very partial list of physical
applications of GA\ methods includes the fields of gravity \cite{gull,
francis}, classical electrodynamics \cite{baylis}, and massive classical
electrodynamics with Dirac's magnetic monopoles \cite{cafaro, cafaro-ali}. In
this paper, however, we are interested in the use of GA in quantum information
and quantum computation. The inclusion of Clifford algebra and GA in quantum
information science (QIS) is quite reasonable, given some physically motivated
grounds \cite{vlasov, doran1}. Indeed, any quantum bit $\left\vert
q\right\rangle $, viewed as the fundamental messenger of quantum information
and realized in terms of a spin-$1/2$\ system, can be considered as a
$2\times\ 2$\ matrix when described in terms of its corresponding density
matrix $\rho=\left\vert q\right\rangle \left\langle q\right\vert $. Once this
link with $2\times2$ matrices is made, one recalls that any $2\times2$ matrix
can be expressed as a combination of Pauli matrices. These, in turn, represent
some GA. Furthermore, any $2\times2$ unitary transformation as well can be
specified by elements of some GA. Motivated by these considerations, the
multiparticle geometric algebra formalism was originally employed in Ref.
\cite{somaroo} to provide a first GA-based reformulation of some of the most
important operations in quantum computation. In a more unconventional use of
GA\ in quantum computing, suitable GA structures were used to perform
quantum-like algorithms without closely considering quantum theory
\cite{marek1, marek2, marek3}. Within this less orthodox approach, the
geometric product replaces the standard tensor product. Moreover, multivectors
interpreted in a geometric fashion by means of \emph{bags of shapes }are used
to specify ordinary quantum entangled states. The GA approach to quantum
computing with states, gates, and quantum algorithms as proposed in Refs.
\cite{marek1, marek2, marek3} generates novel conceptual elements in the field
of QIS. For instance, the microscopic flavor of the quantum computing
formalism is lost when described from the point of view of such a GA approach.
This loss leads to a sort of non-microworld implementation of quantum
computation. This type of implementation is supported by the thesis carried
out in Refs. \cite{marek1,marek2,marek3,caz09,caz20,caz20B}, where it is
stated that there is no fundamentally basic reason why one should assume that
quantum computation must be necessarily associated with physical systems
characterized the rules of quantum theory \cite{marek3}. An extensive
technical discussion on the application of GA\ methods to QIS is presented in
Ref. \cite{doran1}. However, this discussion lacks a presentation on the
fundamentally relevant notion of universality in quantum computing. Despite
that fact that the Toffoli and Fredkin three-qubit quantum gates were
formulated in terms of GA in Ref. \cite{somaroo}, the authors did not consider
any explicit characterization of one and two-qubit quantum gates. In Ref.
\cite{mancio11}, Cafaro and Mancini not only presented an explicit GA
characterization of one-qubit and two-qubit quantum states together with a GA
description of universal sets of quantum gates for quantum computation, they
also demonstrated the universality of a specific set of quantum gates in terms
of the geometric algebra language.

In this paper, given the recent increasing visibility obtained by our findings
reported in Ref. \cite{mancio11} as evident from Refs.
\cite{hild20,ed21,alves22,hild22,J22,silva23,ery23,winter23,J23,J24,toffano24}%
, we present a critical revisitation of our work in Ref. \cite{mancio11}. Our
overall scope is to emphasize the concepts of entanglement and universality in
QIS after offering an instructive GA characterization of both one- and
two-qubit quantum states and quantum gates. More specifically, we begin with
the essential elements of the multiparticle spacetime algebra (MSTA, the
geometric Clifford algebra of a relativistic configuration space
\cite{lasenby93, cd93, doran96, somaroo99}). We then use the MSTA to describe
one- and two-qubit quantum states including, for instance, the $2$-qubit Bell
states that represent maximally entangled quantum states of two qubits. We
then extend the application of the MSTA to specify both one-qubit gates (i.e.,
bit-flip, phase-flip, combined bit and phase flip quantum gates, Hadamard
gate, rotation gate, phase gate and $\pi/8$-gate) and two-qubit quantum
computational gates (i.e., CNOT, controlled-phase and SWAP quantum gates)
\cite{NIELSEN}. Then, employing this proposed GA description of states and
gates along with the GA characterization of the Lie algebras $\mathrm{SO}%
\left(  3\right)  $ and $\mathrm{SU}\left(  2\right)  $ in terms of the rotor
group $\mathrm{Spin}^{+}\left(  3\text{, }0\right)  $ formalism, we revisit
from a GA perspective the proof of universality of quantum gates as discussed
by Boykin and collaborators in Refs. \cite{B99, B00}.

Inspired by Ref. \cite{mancio11}, we made a serious effort here to write this
work with a more pedagogical scope for a wider audience. For this reason, we
added visual schematic depictions together with background GA preliminary
technical details (including, for instance, what appears in Appendix A).
Moreover, we were able to highlight most of the very interesting works we have
partially inspired throughout these years
\cite{hild20,ed21,alves22,hild22,J22,silva23,ery23,winter23,J23,J24,toffano24}%
. Finally, we were able to suggest limitations and, at the same time,
proposals for future research directions compatible with the current
scientific knowledge at the boundaries between GA\ and quantum computing (with
the concept of entanglement playing a prominent role).

The rest of the paper is formally organized as follows. In Section II, we
display the essential ingredients of the MSTA formalism necessary to
characterize quantum states and elementary gates in quantum computing from a
GA viewpoint. In Section III, we offer an explicit GA description of one- and
two-qubit quantum states together with a GA representation of one- and
two-qubit quantum computational gates. In addition, we concisely discuss in
Section III the extension of the MSTA formalism to density matrices for mixed
quantum states. In Section IV, we revisit the proof of universality of quantum
gates as originally provided by Boykin and collaborators in Refs. \cite{B99,
B00} by making use of the material presented in Sections II and III and, in
addition, by exploiting the above-mentioned GA description of the Lie
algebras\textrm{ }$\mathrm{SO}\left(  3\right)  $ and $\mathrm{SU}\left(
2\right)  $ in terms of the rotor group $\mathrm{Spin}^{+}\left(  3\text{,
}0\right)  $ formalism. We present our concluding remarks in Section V.
Finally, some technical details on the algebra of physical space
$\mathfrak{cl}(3)$ and the spacetime Clifford algebra $\mathfrak{cl}(1,3)$
appear in Appendix A.

\section{Basics of Multiparticle Spacetime Algebra}

In this section, we describe the essentials of the MSTA formalism that is
necessary to characterize, from a GA perspective, elementary gates in quantum computing.

From an historical standpoint, GA\ methods were originally introduced in
quantum mechanics via Hestenes' work on understanding the nature of the
electroweak group \cite{hestenes82} and the concept of zitterbewegung within
the spacetime algebra formulation of the Dirac relativistic theory of the
electron \cite{hestenes90}. While these first GA\ explorations into the
quantum world were motivated by the need of seeking deeper insights into
quantum theory, later inspections were motivated by pursuing more practical
computational advantages. For instance, the computational power of GA
techniques in the form of spacetime algebra in quantum mechanics was compared
with the more cumbersome calculations based on explicit matrix formulations in
Ref. \cite{lewis01}. The theoretical characterization of quantum states and
operators from a quantum computing perspective was originally discussed in
Ref. \cite{cd93}. These GA\ characterizations of relevance in quantum
information processing found their first practical applications, for instance,
on the use of quantum gates in NMR (nuclear magnetic resonance) experiments in
Refs. \cite{somaroo99,havel01}.

In the orthodox context for quantum mechanics, it is usually assumed that the
notions of complex space and imaginary unit $i_{%
\mathbb{C}
}$ are essential. Interestingly, employing the geometric Clifford algebra of
real $4$-dimensional Minkowski spacetime \cite{dl} (i.e., the so-called
spacetime algebra (STA)), it can be shown that the $i_{%
\mathbb{C}
}$ that appears in the Dirac, Pauli and Schr\"{o}dinger equations possesses a
clear interpretation in terms of rotations in real spacetime \cite{david75}.
This bouncing between complex and real quantities in quantum mechanics can can
be clearly understood once one introduces the so-called multiparticle
spacetime algebra (MSTA), that is to say, the geometric algebra of a
relativistic configuration space \cite{lasenby93, cd93, doran96, somaroo99}.
In the traditional description of quantum mechanics, tensor products are
employed to construct multiparticle states as well as many of the operators
acting on the states themselves. A tensor product is a formal tool for setting
apart the Hilbert spaces of different particles in an explicit fashion. GA
seeks to explain, from a fundamental viewpoint, the application of the tensor
product in nonrelativistic quantum mechanics by means of the underlying
space-time geometry \cite{cd93}. Within the GA formalism, the\textbf{\ }%
\emph{geometric product }provides a different characterization of the tensor
product. Inspired by the effectiveness of the STA\ formalism in characterizing
a single-particle quantum mechanics, the MSTA\ perspective on multiparticle
quantum mechanics in non-relativistic as well as relativistic settings was
initially proposed with the expectation that it would also deliver
computational and, most of all, interpretational improvements in multiparticle
quantum theory \cite{cd93}. Conceptual advances are expected to emerge thanks
to the peculiar geometric insights furnished by the MSTA formalism. A
distinctive aspect of the MSTA\ is that it requires, for each particle, the
existence of a separate copy of both the time dimension and the three spatial
dimensions. The MSTA\ formalism represents a serious attempt to construct a
convincing conceptual setting for a multi-time perspective on quantum theory.
In conclusion, the primary justification for employing this MSTA formalism is
the attempt of enhancing our comprehension of the very important notions of
\emph{locality} and \emph{causality} in quantum theory \cite{carloIJTP24}.
Indeed, exciting utilizations of the MSTA formalism for the revisitation of
Holland's causal interpretation of a system of two spin-$1/2$\ particles
\cite{holland} are proposed in Refs. \cite{doran96,somaroo99}. In Ref.
\cite{holland}, Holland considers a \ Bell inequality type experiment where
spin measurements are performed on a composite quantum system of two
correlated spin-$1/2$\ particles. In particular, Holland proposes a
non-relativistic definition of local observables that act on the space of a
two-particle wave function and that are extracted from the two-particle wave
function itself. From a conceptual standpoint, one of the limitations of
Holland's analysis is its nonrelativistic nature. Indeed, the notions of
causality and superluminal propagation would require a relativistic setting to
be addressed in a coherent fashion. This was pointed out in Ref.
\cite{doran96}, where a first illustration of the utility of the multiparticle
STA in providing an alternative characterization of the non-locality revealed
by Einstein-Podolsky-Rosen-type experiments in the framework of
nonrelativistic quantum theory was offered. However, the power of the MSTA was
not fully exploited in Ref. \cite{doran96} since the treatment was
nonrelativistic as well. From a computational standpoint, Holland's approach
is based on building a set of tensor variables from quadratic combinations of
the spinorial wave function. Then, compared to the underlying spinorial
degrees of freedom, these tensor variables are shown to be more easily
associated with a set of physical properties. In Ref. \cite{somaroo99}, it was
shown that the MSTA formulation of multiparticle quantum theory makes simpler
the objective of extracting these physical variables considered by Holland in
a considerable way. In this context, the list of advantages that the
GA\ language offers includes the simplification of calculations thanks of the
lack of unessential mathematical technicalities and, in addition, the
clarification of the link between spinorial and tensorial degrees of freedom.
For further details, we suggest Ref. \cite{somaroo99}. Inspired by this line
of research, we apply here the MSTA formalism to describe qubits, quantum
gates, and to revisit the proof of universality in quantum computing as
provided by Boykin and collaborators from a GA viewpoint. For some basic
details on the algebra of physical space $\mathfrak{cl}(3)$ and the spacetime
Clifford algebra $\mathfrak{cl}(1,3)$, we refer to Appendix A. In what
follows, we begin with the $n$-qubit spacetime algebra formalism.

\subsection{The $n$-Qubit Spacetime Algebra Formalism}

The multiparticle spacetime algebra offers an ideal algebraic structure for
the characterization of multiparticle states as well as operators acting on
them. The MSTA is the geometric algebra of $n$-particle configuration space.
\ In particular, for relativistic systems, the $n$-particle configuration
space is composed of $n$-copies of Minkowski spacetime with each copy being a
$1$-particle space. A convenient basis for the MSTA is specified by the set
$\left\{  \gamma_{\mu}^{a}\right\}  $, with $\mu=0$,.., $3$ and $a=1$,.., $n$
identifying the spacetime vector and the particle space, respectively. These
basis vectors $\left\{  \gamma_{\mu}^{a}\right\}  $ fulfill the orthogonality
relations $\gamma_{\mu}^{a}\cdot\gamma_{\nu}^{b}=\delta^{ab}\eta_{\mu\nu}$
where $\eta_{\mu\nu}\overset{\text{def}}{=}$ \textrm{diag}$\left(  +\text{,
}-\text{, }-\text{, }-\right)  $. Observe that, because of the orthogonality
conditions, vectors from different particle spaces anticommute. Furthermore,
since $\dim_{%
\mathbb{R}
}\left[  \mathfrak{cl}\left(  1\text{, }3\right)  \right]  ^{n}=2^{4n}$, a
basis for the entire MSTA possesses $2^{4n}$ degrees of freedom. In the
framework of nonrelativistic quantum mechanics, a single absolute time is used
to identify all of the individual time coordinates. One can pick this vector
to be $\gamma_{0}^{a}$ for each $a$. Then, bivectors are used for modelling
spatial vectors relative to these timelike vectors $\left\{  \gamma_{0}%
^{a}\right\}  $ through a spacetime split. Moreover, a basis set of relative
vectors is specified by $\left\{  \sigma_{k}^{a}\right\}  $ where $\sigma
_{k}^{a}\overset{\text{def}}{=}\gamma_{k}^{a}\gamma_{0}^{a}$ with $k=1$,..,
$3$ and $a=1$,.., $n$. The basis set $\left\{  \sigma_{k}^{a}\right\}  $ gives
rise, for each particle space, to the GA of relative space $\mathfrak{cl}%
(3)\cong\mathfrak{cl}^{+}(1$, $3)$. Each particle space possesses a basis set
specified by,%
\begin{equation}
1\text{, }\left\{  \sigma_{k}\right\}  \text{, }\left\{  i\sigma_{k}\right\}
\text{, }i\text{,} \label{bf}%
\end{equation}
where the volume element $i$ denotes the highest grade multivector known as
the \emph{pseudoscalar}. Neglecting the particle space indices, the
pseudoscalar is defined as $i\overset{\text{def}}{=}\sigma_{1}\sigma_{2}%
\sigma_{3}$. The basis set in Eq. (\ref{bf}) characterizes the Pauli algebra,
the geometric algebra of the $3$-dimensional Euclidean space \cite{dl}.
However, the three Pauli $\sigma_{k}$ are regarded in GA\ as three independent
basis vectors for real space. They are no longer considered as three
matrix-valued components of a single isospace vector. Unlike spacetime basis
vectors, note that $\sigma_{k}^{a}\sigma_{j}^{b}=\sigma_{j}^{b}\sigma_{k}^{a}$
for any $a\neq b$. In other words, relative vectors $\left\{  \sigma_{k}%
^{a}\right\}  $ originating from distinct particle spaces commute. Observe
that the set $\left\{  \sigma_{k}^{a}\right\}  $ give rise to the space
$\left[  \mathfrak{cl}(3)\right]  ^{n}\overset{\text{def}}{=}\mathfrak{cl}%
(3)\otimes$...$\otimes\mathfrak{cl}(3)$ defined as the direct product space of
$n$ copies of $\mathfrak{cl}(3)$, the geometric algebra of the $3$-dimensional
Euclidean space. In the context of the MSTA\ formalism, Pauli spinors can be
viewed as elements of the even subalgebra of the Pauli algebra spanned by
$\left\{  1\text{, }i\sigma_{k}\right\}  $ which, in turn, is isomorphic to
the quaternion algebra. This even subalgebra of the Pauli algebra is a
$4$-dimensional real space in which an arbitrary even element can be recast as
$\psi=a^{0}+a^{k}i\sigma_{k}$, where $a^{0}$ and $a^{k}$ are \emph{real}
scalars for any $k=1$, $2$, $3$. A quantum state in ordinary quantum mechanics
can be specified by a pair of \emph{complex}\textit{ }numbers $\alpha$ and
$\beta$ as,%
\begin{equation}
\left\vert \psi\right\rangle =\left(
\begin{array}
[c]{c}%
\alpha\\
\beta
\end{array}
\right)  =\left(
\begin{array}
[c]{c}%
\operatorname{Re}\alpha+i_{%
\mathbb{C}
}\operatorname{Im}\alpha\\
\operatorname{Re}\beta+i_{%
\mathbb{C}
}\operatorname{Im}\beta
\end{array}
\right)  \text{.}%
\end{equation}
Interestingly, a $1\leftrightarrow1$ map between Pauli column spinors
$\left\{  \left\vert \psi\right\rangle \right\}  $ and elements $\left\{
\psi\right\}  $ of the even subalgebra was shown to be true in Ref.
\cite{lasenby93}. Indeed, one has%
\begin{equation}
\left\vert \psi\right\rangle =\left(
\begin{array}
[c]{c}%
a^{0}+i_{%
\mathbb{C}
}a^{3}\\
-a^{2}+i_{%
\mathbb{C}
}a^{1}%
\end{array}
\right)  \leftrightarrow\psi=a^{0}+a^{k}i\sigma_{k}\text{.} \label{55}%
\end{equation}
with $a^{0}$ and $a^{k}$ being real coefficients for any $k=1$, $2$, $3$. The
set of multivectors $\left\{  1\text{, }i\sigma_{1}\text{, }i\sigma_{2}\text{,
}i\sigma_{3}\right\}  $ denotes the set of computational basis states for the
real $4$-dimensional even subalgebra that corresponds to the two-dimensional
complex Hilbert space $\mathcal{H}_{2}^{1}$ with standard computational basis
specified by $\mathcal{B}_{\mathcal{H}_{2}^{1}}\overset{\text{def}}{=}\left\{
\left\vert 0\right\rangle \text{, }\left\vert 1\right\rangle \right\}  $. In
the context of the GA formalism, the following identifications hold true%
\begin{equation}
\left\vert 0\right\rangle \leftrightarrow\psi_{\left\vert 0\right\rangle
}^{\left(  \text{\textrm{GA}}\right)  }\overset{\text{def}}{=}1\text{, and
}\left\vert 1\right\rangle \leftrightarrow\psi_{\left\vert 1\right\rangle
}^{\left(  \text{\textrm{GA}}\right)  }\overset{\text{def}}{=}-i\sigma
_{2}\text{.}%
\end{equation}
Moreover, in GA\ terms \cite{lasenby93}, the action of the usual quantum Pauli
operators $\left\{  \hat{\Sigma}_{k}\text{, }i_{%
\mathbb{C}
}\hat{I}\right\}  $ on the states $\left\vert \psi\right\rangle $ becomes
\begin{equation}
\hat{\Sigma}_{k}\left\vert \psi\right\rangle \leftrightarrow\sigma_{k}%
\psi\sigma_{3}\text{, } \label{16}%
\end{equation}
with $k=1$, $2$, $3$ and,%
\begin{equation}
i_{%
\mathbb{C}
}\left\vert \psi\right\rangle \leftrightarrow\psi i\sigma_{3}\text{.}%
\end{equation}
Note that $\hat{I}$ is the identity operator on $\mathcal{H}_{2}^{1}$. In
summary, in the framework of the single-particle theory, non-relativistic
states are specified by the even subalgebra of the Pauli algebra with a basis
defined in terms of the set of multivectors $\left\{  1\text{, }i\sigma
_{k}\right\}  $ with $k=1$, $2$, $3$. In particular, right multiplication by
$i\sigma_{3}$ plays the role of the multiplication by the (unique) complex
imaginary unit $i_{%
\mathbb{C}
}$ in ordinary quantum mechanics. Simple calculations suffice to verify that
this translation scheme works in a proper fashion. Indeed, from Eqs.
(\ref{55}) and (\ref{16}), we get%
\begin{align}
\hat{\Sigma}_{1}\left\vert \psi\right\rangle  &  =\left(
\begin{array}
[c]{c}%
-a^{2}+i_{%
\mathbb{C}
}a^{1}\\
a^{0}+i_{%
\mathbb{C}
}a^{3}%
\end{array}
\right)  \leftrightarrow-a^{2}+a^{3}i\sigma_{1}-a^{0}i\sigma_{2}+a^{1}%
i\sigma_{3}=\sigma_{1}\left(  a^{0}+a^{k}i\sigma_{k}\right)  \sigma
_{3}\text{,}\nonumber\\
& \nonumber\\
\hat{\Sigma}_{2}\left\vert \psi\right\rangle  &  =\left(
\begin{array}
[c]{c}%
a^{1}+i_{%
\mathbb{C}
}a^{2}\\
-a^{3}+i_{%
\mathbb{C}
}a^{0}%
\end{array}
\right)  \leftrightarrow a^{1}+a^{0}i\sigma_{1}+a^{3}i\sigma_{2}+a^{2}%
i\sigma_{3}=\sigma_{2}\left(  a^{0}+a^{k}i\sigma_{k}\right)  \sigma
_{3}\text{,}\nonumber\\
& \nonumber\\
\hat{\Sigma}_{3}\left\vert \psi\right\rangle  &  =\left(
\begin{array}
[c]{c}%
a^{0}+i_{%
\mathbb{C}
}a^{3}\\
a^{2}-i_{%
\mathbb{C}
}a^{1}%
\end{array}
\right)  \leftrightarrow a^{0}-a^{1}i\sigma_{1}-a^{2}i\sigma_{2}+a^{3}%
i\sigma_{3}=\sigma_{3}\left(  a^{0}+a^{k}i\sigma_{k}\right)  \sigma
_{3}\text{.}%
\end{align}
It is important to note that, although there are $n$-copies of $i\sigma_{3}$
in the $n$-particle algebra specified by $i\sigma_{3}^{a}$ with $a=1$,.., $n$,
the right-multiplication by all of these $\left\{  i\sigma_{3}^{a}\right\}  $
must yield the same result. This is required to correctly reproduce ordinary
quantum mechanics. For this reason, the following constraints must be imposed%
\begin{equation}
\psi i\sigma_{3}^{1}=\psi i\sigma_{3}^{2}=\text{...}=\psi i\sigma_{3}%
^{n-1}=\psi i\sigma_{3}^{n}\text{.} \label{ccn}%
\end{equation}
The conditions in Eq. (\ref{ccn}) can be obtained by presenting the
$n$-particle correlator $E_{n}$,%
\begin{equation}
E_{n}\overset{\text{def}}{=}%
{\displaystyle\prod\limits_{b=2}^{n}}
\frac{1}{2}\left(  1-i\sigma_{3}^{1}i\sigma_{3}^{b}\right)  \text{,}
\label{en}%
\end{equation}
satisfying the relations $E_{n}i\sigma_{3}^{a}=E_{n}i\sigma_{3}^{b}=J_{n}$ for
any $a$ and $b$. What is $J_{n}?$ Observe that $E_{n}$ in Eq. (\ref{en}) has
been introduced by selecting the $a=1$ space and, then, correlating all the
other spaces to this space. However, the value of $E_{n}$ does not depend on
which one of the $n$ spaces is picked and correlated to. The complex structure
is characterized by $J_{n}\overset{\text{def}}{=}E_{n}i\sigma_{3}^{a}$, where
$J_{n}^{2}=-E_{n}$. One can notice that the number of \emph{real} degrees of
freedom are reduced from $4^{n}=\dim_{%
\mathbb{R}
}\left[  \mathfrak{cl}^{+}(3)\right]  ^{n}$ to the expected $2^{n+1}=\dim_{%
\mathbb{R}
}\mathcal{H}_{2}^{n}$ thanks to the right-multiplication by the quantum
correlator $E_{n}$, which, in turn, can be regarded as acting as a projection
operator. From a physical standpoint, the projection locks the phases of the
various particles together. The \emph{reduced} even subalgebra space is
generally denoted by $\left[  \mathfrak{cl}^{+}(3)\right]  ^{n}/E_{n}$. Then,
in analogy to $\mathfrak{cl}^{+}(3)$ for a single particle, multivectors that
belong to $\left[  \mathfrak{cl}^{+}(3)\right]  ^{n}/E_{n}$ can be viewed as
$n$-particle spinors (or, alternatively, $n$-qubit states). Summing up, the
generalization to multiparticle systems requires, for each particle, a
separate copy of the STA. Moreover, the usual complex imaginary unit induces
correlations between these particle spaces.

Although we presented a general GA framework for $n$-particles\ quantum theory
in terms of a relativistic spacetime algebra, many of the necessary properties
can be illustrated by focusing on two-particle systems. Interestingly, both
classical relativistic physics and the standard quantum formalism have a
spinorial formulation in the GA language. The algebraic employment of spinors,
in particular, offers quantum-mechanical character to several classical
findings. For more details, we refer to \cite{sob04}. In what follows, indeed,
we focus on the special case of the $2$-qubit spacetime algebra formalism.

\subsection{The $2$-Qubit Spacetime Algebra Formalism}

As previously mentioned, while quantum mechanics has a unique imaginary unit
$i_{%
\mathbb{C}
}$, the $2$-particle algebra possesses two bivectors playing the role of $i_{%
\mathbb{C}
}$, namely $i\sigma_{3}^{1}$ and $i\sigma_{3}^{2}$. To properly reproduce
ordinary quantum mechanics, right-multiplication of a state by either of these
bivectors must yield the same state. Therefore, it is mandatory that we have%
\begin{equation}
\psi i\sigma_{3}^{1}=\psi i\sigma_{3}^{2}\text{.} \label{a}%
\end{equation}
Manipulating Eq. (\ref{a}) leads to $\psi=\psi E$, with $E=E^{2}$ specified
by
\begin{equation}
E\overset{\text{def}}{=}\frac{1}{2}\left(  1-i\sigma_{3}^{1}i\sigma_{3}%
^{2}\right)  \text{.} \label{EE}%
\end{equation}
Following what we stated in the previous subsection, right-multiplication by
$E$ is a projection operation. Moreover, the number of \emph{real }degrees of
freedom drops from $16$ to the expected $8$ once we include this factor $E$ on
the right-hand-side of all states. The $16$-dimensional geometric algebra
$\mathfrak{cl}^{+}(3)\otimes\mathfrak{cl}^{+}(3)$ can be spanned by the set
multivectors that specify the basis $\mathcal{B}_{\mathfrak{cl}^{+}%
(3)\otimes\mathfrak{cl}^{+}(3)}$ defined as,%
\begin{equation}
\mathcal{B}_{\mathfrak{cl}^{+}(3)\otimes\mathfrak{cl}^{+}(3)}\overset
{\text{def}}{=}\left\{  1\text{, }i\sigma_{l}^{1}\text{, }i\sigma_{k}%
^{2}\text{, }i\sigma_{l}^{1}i\sigma_{k}^{2}\text{ }\right\}  \text{,}
\label{ci1}%
\end{equation}
with $k,l=1$, $2$, $3$. Using the quantum projection operator $E$ to
right-multiply the multivectors in $\mathcal{B}_{\mathfrak{cl}^{+}%
(3)\otimes\mathfrak{cl}^{+}(3)}$, we get%
\begin{equation}
\mathcal{B}_{\mathfrak{cl}^{+}(3)\otimes\mathfrak{cl}^{+}(3)}\overset
{E}{\rightarrow}\mathcal{B}_{\mathfrak{cl}^{+}(3)\otimes\mathfrak{cl}^{+}%
(3)}E\overset{\text{def}}{=}\left\{  E\text{, }i\sigma_{l}^{1}E\text{,
}i\sigma_{k}^{2}E\text{, }i\sigma_{l}^{1}i\sigma_{k}^{2}E\text{ }\right\}
\text{.}%
\end{equation}
After some simple calculations, one finds that%
\begin{align}
E  &  =-i\sigma_{3}^{1}i\sigma_{3}^{2}E\text{, }i\sigma_{1}^{2}E=-i\sigma
_{3}^{1}i\sigma_{2}^{2}E\text{, }i\sigma_{2}^{2}E=i\sigma_{3}^{1}i\sigma
_{1}^{2}E\text{, }i\sigma_{3}^{2}E=i\sigma_{3}^{1}E\text{,}\nonumber\\
& \nonumber\\
\text{ }i\sigma_{1}^{1}E  &  =-i\sigma_{2}^{1}i\sigma_{3}^{2}E\text{, }%
i\sigma_{1}^{1}i\sigma_{1}^{2}E=-i\sigma_{2}^{1}i\sigma_{2}^{2}E\text{,
}i\sigma_{1}^{1}i\sigma_{2}^{2}E=i\sigma_{2}^{1}i\sigma_{1}^{2}E\text{,
}i\sigma_{1}^{1}i\sigma_{3}^{2}E=i\sigma_{2}^{1}E\text{.} \label{ci2}%
\end{align}
Therefore, using Eqs. (\ref{ci1}) and (\ref{ci2}), a convenient basis for the
$8$-dimensional \emph{reduced} even subalgebra $\left[  \mathfrak{cl}%
^{+}(3)\otimes\mathfrak{cl}^{+}(3)\right]  /E$ can be expressed as,%
\begin{equation}
\mathcal{B}_{\left[  \mathfrak{cl}^{+}(3)\otimes\mathfrak{cl}^{+}(3)\right]
/E}\overset{\text{def}}{=}\left\{  1\text{, }i\sigma_{1}^{2}\text{, }%
i\sigma_{2}^{2}\text{, }i\sigma_{3}^{2}\text{, }i\sigma_{1}^{1}\text{,
}i\sigma_{1}^{1}i\sigma_{1}^{2}\text{, }i\sigma_{1}^{1}i\sigma_{2}^{2}\text{,
}i\sigma_{1}^{1}i\sigma_{3}^{2}\right\}  \text{.} \label{bb}%
\end{equation}
The basis $\mathcal{B}_{\left[  \mathfrak{cl}^{+}(3)\otimes\mathfrak{cl}%
^{+}(3)\right]  /E}$ in Eq. (\ref{bb}) spans $\left[  \mathfrak{cl}%
^{+}(3)\otimes\mathfrak{cl}^{+}(3)\right]  /E$ and corresponds to a proper
ordinary complex basis that generates the complex Hilbert space $\mathcal{H}%
_{2}^{2}$. A $2$-qubits quantum state or, alternatively, a direct-product
$2$-particle Pauli spinor can be represented in the framework of spacetime
algebra in terms of $\psi^{1}\phi^{2}E$, namely $\left\vert \psi\text{, }%
\phi\right\rangle \leftrightarrow\psi^{1}\phi^{2}E$, with $\psi^{1}$ and
$\phi^{2}$ being even multivectors in their own spaces. A GA description of
the usual computational basis for $2$-particle spin states is given by,%
\begin{align}
\left\vert 0\right\rangle \otimes\left\vert 0\right\rangle  &  =\left(
\begin{array}
[c]{c}%
1\\
0
\end{array}
\right)  \otimes\left(
\begin{array}
[c]{c}%
1\\
0
\end{array}
\right)  \leftrightarrow E\text{, }\left\vert 0\right\rangle \otimes\left\vert
1\right\rangle =\left(
\begin{array}
[c]{c}%
1\\
0
\end{array}
\right)  \otimes\left(
\begin{array}
[c]{c}%
0\\
1
\end{array}
\right)  \leftrightarrow-i\sigma_{2}^{2}E\text{,}\nonumber\\
& \nonumber\\
\left\vert 1\right\rangle \otimes\left\vert 0\right\rangle  &  =\left(
\begin{array}
[c]{c}%
0\\
1
\end{array}
\right)  \otimes\left(
\begin{array}
[c]{c}%
1\\
0
\end{array}
\right)  \leftrightarrow-i\sigma_{2}^{1}E\text{, }\left\vert 1\right\rangle
\otimes\left\vert 1\right\rangle =\left(
\begin{array}
[c]{c}%
0\\
1
\end{array}
\right)  \otimes\left(
\begin{array}
[c]{c}%
0\\
1
\end{array}
\right)  \leftrightarrow i\sigma_{2}^{1}i\sigma_{2}^{2}E\text{.} \label{d2}%
\end{align}
In particular, recall that a typical maximally entangled state between a pair
of $2$-level systems can be written as,%
\begin{equation}
\left\vert \psi_{\text{\textrm{singlet}}}\right\rangle \overset{\text{def}}%
{=}\frac{1}{\sqrt{2}}\left\{  \left(
\begin{array}
[c]{c}%
1\\
0
\end{array}
\right)  \otimes\left(
\begin{array}
[c]{c}%
0\\
1
\end{array}
\right)  -\left(
\begin{array}
[c]{c}%
0\\
1
\end{array}
\right)  \otimes\left(
\begin{array}
[c]{c}%
1\\
0
\end{array}
\right)  \right\}  =\frac{1}{\sqrt{2}}\left(  \left\vert 01\right\rangle
-\left\vert 10\right\rangle \right)  \text{.} \label{d3}%
\end{equation}
Using Eqs. (\ref{EE}), (\ref{d2}) and (\ref{d3}), the GA version of
$\left\vert \psi_{\text{singlet}}\right\rangle $ in Eq. (\ref{d3}) is given
by,%
\begin{equation}
\mathcal{H}_{2}^{2}\ni\left\vert \psi_{\text{\textrm{singlet}}}\right\rangle
\leftrightarrow\psi_{\text{\textrm{singlet}}}^{\left(  \text{\textrm{GA}%
}\right)  }\in\left[  \mathfrak{cl}^{+}(3)\right]  ^{2}\text{,}%
\end{equation}
with $\psi_{\text{\textrm{singlet}}}^{\left(  \text{\textrm{GA}}\right)  }$
equal to,%
\begin{equation}
\psi_{\text{\textrm{singlet}}}^{\left(  \text{\textrm{GA}}\right)  }=\frac
{1}{2^{\frac{3}{2}}}\left(  i\sigma_{2}^{1}-i\sigma_{2}^{2}\right)  \left(
1-i\sigma_{3}^{1}i\sigma_{3}^{2}\right)  \text{.}%
\end{equation}
Moreover, the right-sided multiplication by $J$ replaces the the role of
multiplication by the complex imaginary unit $i_{%
\mathbb{C}
}$ for $2$-particle spin states,%
\begin{equation}
J=Ei\sigma_{3}^{1}=Ei\sigma_{3}^{2}=\frac{1}{2}\left(  i\sigma_{3}^{1}%
+i\sigma_{3}^{2}\right)  \text{,}%
\end{equation}
in such a manner that $J^{2}=-E$ with $E$ in Eq. (\ref{EE}). From a
GA\ perspective, the action on $2$-particle spin states of $2$-particle Pauli
operators is specified by%
\begin{equation}
\hat{\Sigma}_{k}\otimes\hat{I}\left\vert \psi\right\rangle \leftrightarrow
-i\sigma_{k}^{1}\psi J\text{, }\hat{\Sigma}_{k}\otimes\hat{\Sigma}%
_{l}\left\vert \psi\right\rangle \leftrightarrow-i\sigma_{k}^{1}i\sigma
_{l}^{2}\psi E\text{, }\hat{I}\otimes\hat{\Sigma}_{k}\left\vert \psi
\right\rangle \leftrightarrow-i\sigma_{k}^{2}\psi J\text{.} \label{impo1}%
\end{equation}
For illustrative purposes, the second correspondence in Eq. (\ref{impo1})
emerges as follows,%
\begin{equation}
\hat{\Sigma}_{l}^{2}\left\vert \psi\right\rangle \leftrightarrow\sigma_{l}%
^{2}\psi\sigma_{3}^{2}=\sigma_{l}^{2}\psi E\sigma_{3}^{2}=-\sigma_{l}^{2}\psi
Eii\sigma_{3}^{2}=-i\sigma_{l}^{2}\psi Ei\sigma_{3}^{2}=-i\sigma_{l}^{2}\psi
J\text{,}%
\end{equation}
and, as a consequence,%
\begin{equation}
\hat{\Sigma}_{k}\otimes\hat{\Sigma}_{l}\left\vert \psi\right\rangle
\leftrightarrow\left(  -i\sigma_{k}^{1}\right)  \left(  -i\sigma_{l}%
^{2}\right)  \psi J^{2}=-i\sigma_{k}^{1}i\sigma_{l}^{2}\psi E\text{.}%
\end{equation}
Finally, recollecting that $i_{%
\mathbb{C}
}\hat{\Sigma}_{k}\left\vert \psi\right\rangle \leftrightarrow i\sigma_{k}\psi
$, we emphasize that%
\begin{equation}
i_{%
\mathbb{C}
}\hat{\Sigma}_{k}\otimes\hat{I}\left\vert \psi\right\rangle \leftrightarrow
i\sigma_{k}^{1}\psi\text{ and, }\hat{I}\otimes i_{%
\mathbb{C}
}\hat{\Sigma}_{k}\left\vert \psi\right\rangle \leftrightarrow i\sigma_{k}%
^{2}\psi\text{.}%
\end{equation}
For additional technical details on the MSTA\ formalism, we suggest Refs.
\cite{lasenby93, cd93, doran96, somaroo99}.

Before moving to the next section, we add here a comment on entanglement and
GA. It is known in quantum theory that when two subsystems of a larger
composite quantum system interact, they become entangled. Then, each one of
these subsystems cannot be characterized by a pure quantum state. When the
total number of subsystems is just two, the Schmidt decomposition method can
be used to quantify the degree of entanglement that appears in the composite
system \cite{NIELSEN}. However, quantifying quantum entanglement in composite
quantum systems that contain more than two subsystems is much more complicated
than characterizing entanglement in bipartite systems. Indeed, even focusing
on pure states, the transition from two to three subsystems exhibits tangible
complications. For instance, while the entanglement properties for an
arbitrary pure state of two subsystems with $d$-levels each can be fully
described by its Schmidt vector, the same is not possible for an arbitrary
tripartite pure state. For a detailed discussion on the crucial differences
between bipartite and multipartite settings in the study of quantum
entanglement, we indicate Refs. \cite{eisert01,plenio07,karol17}.\textbf{
}While the GA approach does not offer a definitive solution to how to quantify
the entanglement degree of multipartite quantum systems, it offers the
advantage that the number of entangled particles only modifies the size of the
space one is working in. However, it does not change the type of entanglement
analysis employed when transitioning from two-subsystems to $n$-subsystems
with $n>2$. For an in depth discussion on the GA form of the Schmidt
decomposition and its possible extension to quantifying multipartite
entanglement, we refer to Ref. \cite{parker01}.

\section{Quantum Computing with Geometric Algebra}

In general, nontrivial quantum computations that occur in quantum algorithms
can demand the construction of tricky computational networks characterized by
a large number of gates that act on $n$-qubit quantum states. For this reason,
it is very important to find a suitable \emph{universal} set of quantum gates.
From a formal standpoint, a set of quantum gates $\left\{  \hat{U}%
_{i}\right\}  $ is considered to be \textit{universal} if any logical
operation $\hat{U}_{L}$ can be decomposed as \cite{NIELSEN} ,%
\begin{equation}
\hat{U}_{L}=%
{\displaystyle\prod\limits_{\hat{U}_{l}\in\left\{  \hat{U}_{i}\right\}  }}
\hat{U}_{l}\text{.}%
\end{equation}
In what follows, we provide a clear GA characterization of $1$- and $2$-qubit
quantum state, along with a GA description of a universal set of quantum gates
for quantum computing. Finally, we briefly discuss about the generalization of
the MSTA formalism to density matrices for mixed quantum states.

\subsection{1-Qubit Quantum Computing}

We begin by considering, in the GA setting, relatively simple circuit models
of quantum computing with $1$-qubit quantum gates.

$\emph{Quantum}$ $\emph{NOT\ Gate}$ $\emph{(or}$ $\emph{Bit}$ $\emph{Flip}$
$\emph{Quantum}$ $\emph{Gate)}$. The NOT gate is represented here by the
symbol $\hat{\Sigma}_{1}$ and denotes a nontrivial reversible operation that
can be applied to a single qubit. For simplicity, we begin using the GA
formalism to investigate the action of quantum gates on $1$-qubit quantum
states given by $\psi_{\left\vert q\right\rangle }^{\left(  \text{\textrm{GA}%
}\right)  }=a^{0}+a^{2}i\sigma_{2}$. Then, the action of the operator
$\hat{\Sigma}_{1}^{\left(  \text{\textrm{GA}}\right)  }$ in the GA setting is
specified by
\begin{equation}
\hat{\Sigma}_{1}\left\vert q\right\rangle \overset{\text{def}}{=}\left\vert
q\oplus1\right\rangle \leftrightarrow\psi_{\left\vert q\oplus1\right\rangle
}^{\left(  \text{\textrm{GA}}\right)  }\overset{\text{def}}{=}\sigma
_{1}\left(  a^{0}+a^{2}i\sigma_{2}\right)  \sigma_{3}\text{.} \label{billy}%
\end{equation}
Given that the unit pseudoscalar $i\overset{\text{def}}{=}\sigma_{1}\sigma
_{2}\sigma_{3}$ satisfies the conditions $i\sigma_{k}=\sigma_{k}i$ with $k=1$,
$2$, $3$ and, in addition, remembering the geometric product rule,%
\begin{equation}
\sigma_{i}\sigma_{j}=\sigma_{i}\cdot\sigma_{j}+\sigma_{i}\wedge\sigma
_{j}=\delta_{ij}+i\varepsilon_{ijk}\sigma_{k}\text{,} \label{aa}%
\end{equation}
Eq. (\ref{billy}) becomes%
\begin{equation}
\hat{\Sigma}_{1}\left\vert q\right\rangle \overset{\text{def}}{=}\left\vert
q\oplus1\right\rangle \leftrightarrow\psi_{\left\vert q\oplus1\right\rangle
}^{\left(  \text{\textrm{GA}}\right)  }=-\left(  a^{2}+a^{0}i\sigma
_{2}\right)  \text{.} \label{uno}%
\end{equation}
For completeness, we emphasize that the action of the unitary quantum gate
$\hat{\Sigma}_{1}^{\left(  \text{\textrm{GA}}\right)  }$ on the GA
computational basis states $\left\{  1\text{, }i\sigma_{1}\text{, }i\sigma
_{2}\text{, }i\sigma_{3}\right\}  $ is specified by the following relations,%
\begin{equation}
\hat{\Sigma}_{1}^{\left(  \text{\textrm{GA}}\right)  }:1\rightarrow
-i\sigma_{2}\text{, }\hat{\Sigma}_{1}^{\left(  \text{\textrm{GA}}\right)
}:i\sigma_{1}\rightarrow i\sigma_{3}\text{, }\hat{\Sigma}_{1}^{\left(
\text{\textrm{GA}}\right)  }:i\sigma_{2}\rightarrow-1\text{, }\hat{\Sigma}%
_{1}^{\left(  \text{\textrm{GA}}\right)  }:i\sigma_{3}\rightarrow i\sigma
_{1}\text{.}%
\end{equation}

\emph{Phase Flip Quantum Gate. }The phase flip gate is denoted by the symbol
$\hat{\Sigma}_{3}$ and is an example of an additional nontrivial reversible
gate that can be applied to a single qubit. The action of the unitary quantum
gate $\hat{\Sigma}_{3}^{\left(  \text{\textrm{GA}}\right)  }$ on the
multivector $\psi_{\left\vert q\right\rangle }^{\left(  \text{\textrm{GA}%
}\right)  }=a^{0}+a^{2}i\sigma_{2}$ can be specified in GA terms as,%
\begin{equation}
\hat{\Sigma}_{3}\left\vert q\right\rangle \overset{\text{def}}{=}\left(
-1\right)  ^{q}\left\vert q\right\rangle \leftrightarrow\psi_{\left(
-1\right)  ^{q}\left\vert q\right\rangle }^{\left(  \text{\textrm{GA}}\right)
}\overset{\text{def}}{=}\sigma_{3}\left(  a^{0}+a^{2}i\sigma_{2}\right)
\sigma_{3}\text{.}%
\end{equation}
Employing Eqs. (\ref{a}) and (\ref{aa}), it happens that%
\begin{equation}
\hat{\Sigma}_{3}\left\vert q\right\rangle \overset{\text{def}}{=}\left(
-1\right)  ^{q}\left\vert q\right\rangle \leftrightarrow\psi_{\left(
-1\right)  ^{q}\left\vert q\right\rangle }^{\left(  \text{\textrm{GA}}\right)
}=a^{0}-a^{2}i\sigma_{2}\text{.} \label{due}%
\end{equation}
Finally, the action of the unitary quantum gate $\hat{\Sigma}_{3}^{\left(
\text{\textrm{GA}}\right)  }$on the basis states $\left\{  1\text{, }%
i\sigma_{1}\text{, }i\sigma_{2}\text{, }i\sigma_{3}\right\}  $ is given by,%
\begin{equation}
\hat{\Sigma}_{3}^{\left(  \text{\textrm{GA}}\right)  }:1\rightarrow1\text{,
}\hat{\Sigma}_{3}^{\left(  \text{\textrm{GA}}\right)  }:i\sigma_{1}%
\rightarrow-i\sigma_{1}\text{, }\hat{\Sigma}_{3}^{\left(  \text{\textrm{GA}%
}\right)  }:i\sigma_{2}\rightarrow-i\sigma_{2}\text{, }\hat{\Sigma}%
_{3}^{\left(  \text{\textrm{GA}}\right)  }:i\sigma_{3}\rightarrow i\sigma
_{3}\text{.}%
\end{equation}

\emph{Combined Bit and Phase Flip Quantum Gates. }A different example of a
nontrivial reversible operation that can be applied to a single qubit can be
constructed by conveniently combining the above mentioned two reversible
operations $\hat{\Sigma}_{1}$ and $\hat{\Sigma}_{3}$. The symbol for the new
operation is $\hat{\Sigma}_{2}$ $\overset{\text{def}}{=}i_{%
\mathbb{C}
}\hat{\Sigma}_{1}\circ\hat{\Sigma}_{3}$ and its action on $\ \psi_{\left\vert
q\right\rangle }^{\left(  \text{\textrm{GA}}\right)  }=a^{0}+a^{2}i\sigma_{2}$
is specified by,
\begin{equation}
\hat{\Sigma}_{2}\left\vert q\right\rangle \overset{\text{def}}{=}i_{%
\mathbb{C}
}\left(  -1\right)  ^{q}\left\vert q\oplus1\right\rangle \leftrightarrow
\psi_{i_{%
\mathbb{C}
}\left(  -1\right)  ^{q}\left\vert q\oplus1\right\rangle }^{\left(
\text{\textrm{GA}}\right)  }\overset{\text{def}}{=}\sigma_{2}\left(
a^{0}+a^{2}i\sigma_{2}\right)  \sigma_{3}\text{.}%
\end{equation}
Employing Eqs. (\ref{a}) and (\ref{aa}), it happens that%
\begin{equation}
\hat{\Sigma}_{2}\left\vert q\right\rangle \overset{\text{def}}{=}i_{%
\mathbb{C}
}\left(  -1\right)  ^{q}\left\vert q\oplus1\right\rangle \leftrightarrow
\psi_{i_{%
\mathbb{C}
}\left(  -1\right)  ^{q}\left\vert q\oplus1\right\rangle }^{\left(
\text{\textrm{GA}}\right)  }=\left(  a^{2}-a^{0}i\sigma_{2}\right)
i\sigma_{3}\text{.}%
\end{equation}
As a matter of fact, making use of Eq. (\ref{aa}) and, in addition, exploiting
the relations $i\sigma_{k}=\sigma_{k}i$ for $k=1$, $2$, $3$, we get
\begin{equation}
\sigma_{2}\left(  a^{0}+a^{2}i\sigma_{2}\right)  \sigma_{3}=\left(
a^{2}-a^{0}i\sigma_{2}\right)  i\sigma_{3}\text{.}%
\end{equation}
Finally, the unitary quantum gate $\hat{\Sigma}_{2}^{\left(  \text{\textrm{GA}%
}\right)  }$ acts on the basis states $\left\{  1\text{, }i\sigma_{1}\text{,
}i\sigma_{2}\text{, }i\sigma_{3}\right\}  $ as,%
\begin{equation}
\hat{\Sigma}_{2}^{\left(  \text{\textrm{GA}}\right)  }:1\rightarrow
i\sigma_{1}\text{, }\hat{\Sigma}_{2}^{\left(  \text{\textrm{GA}}\right)
}:i\sigma_{1}\rightarrow1\text{, }\hat{\Sigma}_{2}^{\left(  \text{\textrm{GA}%
}\right)  }:i\sigma_{2}\rightarrow i\sigma_{3}\text{, }\hat{\Sigma}%
_{2}^{\left(  \text{\textrm{GA}}\right)  }:i\sigma_{3}\rightarrow i\sigma
_{2}\text{.}%
\end{equation}

\emph{Hadamard Quantum Gate. }The GA quantity that corresponds to the
Walsh-Hadamard quantum gate $\hat{H}$ $\overset{\text{def}}{=}\frac
{\hat{\Sigma}_{1}+\hat{\Sigma}_{3}}{\sqrt{2}}$ is denoted here with $\hat
{H}^{\left(  \text{\textrm{GA}}\right)  }$. Its action on $\psi_{\left\vert
q\right\rangle }^{\left(  \text{\textrm{GA}}\right)  }=a^{0}+a^{2}i\sigma_{2}$
is given by,
\begin{equation}
\hat{H}\left\vert q\right\rangle \overset{\text{def}}{=}\frac{1}{\sqrt{2}%
}\left[  \left\vert q\oplus1\right\rangle +\left(  -1\right)  ^{q}\left\vert
q\right\rangle \right]  \leftrightarrow\psi_{\hat{H}\left\vert q\right\rangle
}^{\left(  \text{\textrm{GA}}\right)  }\overset{\text{def}}{=}\left(
\frac{\sigma_{1}+\sigma_{3}}{\sqrt{2}}\right)  \left(  a^{0}+a^{2}i\sigma
_{2}\right)  \sigma_{3}\text{.} \label{tre}%
\end{equation}
Making use of Eqs. (\ref{uno}) and (\ref{due}), the correspondence in Eq.
(\ref{tre}) reduces to%
\begin{equation}
\hat{H}\left\vert q\right\rangle \overset{\text{def}}{=}\frac{1}{\sqrt{2}%
}\left[  \left\vert q\oplus1\right\rangle +\left(  -1\right)  ^{q}\left\vert
q\right\rangle \right]  \leftrightarrow\psi_{\hat{H}\left\vert q\right\rangle
}^{\left(  \text{\textrm{GA}}\right)  }=\frac{a^{0}}{\sqrt{2}}\left(
1-i\sigma_{2}\right)  -\frac{a^{2}}{\sqrt{2}}\left(  1+i\sigma_{2}\right)
\text{.}%
\end{equation}
As a side remark, observe that the GA\ multivectors that correspond to
$\left\vert +\right\rangle $ and $\left\vert -\right\rangle $ (i.e., the
Hadamard transformed computational states) are described as,
\begin{equation}
\left\vert +\right\rangle \overset{\text{def}}{=}\frac{\left\vert
0\right\rangle +\left\vert 1\right\rangle }{\sqrt{2}}\leftrightarrow
\psi_{\left\vert +\right\rangle }^{\left(  \text{\textrm{GA}}\right)  }%
=\frac{1-i\sigma_{2}}{\sqrt{2}}\text{ and, }\left\vert -\right\rangle
\overset{\text{def}}{=}\frac{\left\vert 0\right\rangle -\left\vert
1\right\rangle }{\sqrt{2}}\leftrightarrow\psi_{\left\vert -\right\rangle
}^{\left(  \text{\textrm{GA}}\right)  }=\frac{1+i\sigma_{2}}{\sqrt{2}}\text{,}%
\end{equation}
respectively. Finally, the action of the unitary quantum gate $\hat
{H}^{\left(  \text{\textrm{GA}}\right)  }$ on the basis states $\left\{
1\text{, }i\sigma_{1}\text{, }i\sigma_{2}\text{, }i\sigma_{3}\right\}  $ is%
\begin{equation}
\hat{H}^{\left(  \text{\textrm{GA}}\right)  }:1\rightarrow\frac{1-i\sigma_{2}%
}{\sqrt{2}}\text{, }\hat{H}^{\left(  \text{\textrm{GA}}\right)  }:i\sigma
_{1}\rightarrow\frac{-i\sigma_{1}+i\sigma_{3}}{\sqrt{2}}\text{, }\hat
{H}^{\left(  \text{\textrm{GA}}\right)  }:i\sigma_{2}\rightarrow
-\frac{1+i\sigma_{2}}{\sqrt{2}}\text{, }\hat{H}^{\left(  \text{\textrm{GA}%
}\right)  }:i\sigma_{3}\rightarrow\frac{i\sigma_{1}+i\sigma_{3}}{\sqrt{2}%
}\text{.}%
\end{equation}

\emph{Rotation Gate. }The rotation gate $\hat{R}_{\theta}^{\left(
\text{\textrm{GA}}\right)  }$ acts on $\psi_{\left\vert q\right\rangle
}^{\left(  \text{\textrm{GA}}\right)  }=a^{0}+a^{2}i\sigma_{2}$ as,%
\begin{equation}
\hat{R}_{\theta}\left\vert q\right\rangle \overset{\text{def}}{=}\left[
\frac{1+\exp\left(  i_{%
\mathbb{C}
}\theta\right)  }{2}+\left(  -1\right)  ^{q}\frac{1-\exp\left(  i_{%
\mathbb{C}
}\theta\right)  }{2}\right]  \left\vert q\right\rangle \leftrightarrow
\psi_{\hat{R}_{\theta}\left\vert q\right\rangle }^{\left(  \text{\textrm{GA}%
}\right)  }\overset{\text{def}}{=}a^{0}+a^{2}i\sigma_{2}\left(  \cos
\theta+i\sigma_{3}\sin\theta\right)  \text{.}%
\end{equation}
More generally, the action of the unitary quantum gate $\hat{R}_{\theta
}^{\left(  \text{\textrm{GA}}\right)  }$ on the basis states $\left\{
1\text{, }i\sigma_{1}\text{, }i\sigma_{2}\text{, }i\sigma_{3}\right\}  $ is
given by,%
\begin{equation}
\hat{R}_{\theta}^{\left(  \text{\textrm{GA}}\right)  }:1\rightarrow1\text{,
}\hat{R}_{\theta}^{\left(  \text{\textrm{GA}}\right)  }:i\sigma_{1}\rightarrow
i\sigma_{1}\left(  \cos\theta+i\sigma_{3}\sin\theta\right)  \text{, }\hat
{R}_{\theta}^{\left(  \text{\textrm{GA}}\right)  }:i\sigma_{2}\rightarrow
i\sigma_{2}\left(  \cos\theta+i\sigma_{3}\sin\theta\right)  \text{, }\hat
{R}_{\theta}^{\left(  \text{\textrm{GA}}\right)  }:i\sigma_{3}\rightarrow
i\sigma_{3}\text{.}%
\end{equation}

\emph{Phase Quantum Gate and }$\pi/8$\emph{-Quantum Gate. }The action of the
phase gate $\hat{S}^{\left(  \text{\textrm{GA}}\right)  }$ on $\psi
_{\left\vert q\right\rangle }^{\left(  \text{\textrm{GA}}\right)  }%
=a^{0}+a^{2}i\sigma_{2}$ is given by,%
\begin{equation}
\hat{S}\left\vert q\right\rangle \overset{\text{def}}{=}\left[  \frac{1+i_{%
\mathbb{C}
}}{2}+\left(  -1\right)  ^{q}\frac{1-i_{%
\mathbb{C}
}}{2}\right]  \left\vert q\right\rangle \leftrightarrow\psi_{\hat{S}\left\vert
q\right\rangle }^{\left(  \text{\textrm{GA}}\right)  }\overset{\text{def}}%
{=}a^{0}+\left(  a^{2}i\sigma_{2}\right)  i\sigma_{3}\text{.}%
\end{equation}

Moreover, the action of the unitary quantum gate $\hat{S}^{\left(
\text{\textrm{GA}}\right)  }$ on the basis states $\left\{  1\text{, }%
i\sigma_{1}\text{, }i\sigma_{2}\text{, }i\sigma_{3}\right\}  $ is,%
\begin{equation}
\hat{S}^{\left(  \text{\textrm{GA}}\right)  }:1\rightarrow1\text{, }\hat
{S}^{\left(  \text{\textrm{GA}}\right)  }:i\sigma_{1}\rightarrow i\sigma
_{2}\text{, }\hat{S}^{\left(  \text{\textrm{GA}}\right)  }:i\sigma
_{2}\rightarrow-i\sigma_{1}\text{, }\hat{S}^{\left(  \text{\textrm{GA}%
}\right)  }:i\sigma_{3}\rightarrow i\sigma_{3}\text{.}%
\end{equation}

The GA version of the $\pi/8$-quantum gate $\hat{T}$ is specified by the
following correspondence,%
\begin{equation}
\hat{T}\left\vert q\right\rangle \overset{\text{def}}{=}\left[  \frac
{1+\exp\left(  i_{%
\mathbb{C}
}\frac{\pi}{4}\right)  }{2}+\left(  -1\right)  ^{q}\frac{1-\exp\left(  i_{%
\mathbb{C}
}\frac{\pi}{4}\right)  }{2}\right]  \left\vert q\right\rangle \leftrightarrow
\psi_{\hat{T}\left\vert q\right\rangle }^{\left(  \text{\textrm{GA}}\right)
}\overset{\text{def}}{=}\frac{1}{\sqrt{2}}\left(  a^{0}+a^{2}i\sigma
_{2}\right)  \left(  1+i\sigma_{3}\right)  \text{.}%
\end{equation}

Finally, the action of the unitary quantum gate $\hat{T}^{\left(
\text{\textrm{GA}}\right)  }$ on the basis states $\left\{  1\text{, }%
i\sigma_{1}\text{, }i\sigma_{2}\text{, }i\sigma_{3}\right\}  $ is,%
\begin{equation}
\hat{T}^{\left(  \text{\textrm{GA}}\right)  }:1\rightarrow1\text{, }\hat
{T}^{\left(  \text{\textrm{GA}}\right)  }:i\sigma_{1}\rightarrow i\sigma
_{1}\frac{\left(  1+i\sigma_{3}\right)  }{\sqrt{2}}\text{, }\hat{T}^{\left(
\text{\textrm{GA}}\right)  }:i\sigma_{2}\rightarrow i\sigma_{2}\frac{\left(
1+i\sigma_{3}\right)  }{\sqrt{2}}\text{, }\hat{T}^{\left(  \text{\textrm{GA}%
}\right)  }:i\sigma_{3}\rightarrow i\sigma_{3}\text{.}%
\end{equation}
In Table I, we report the the action of some of the most relevant $1$-qubit
quantum gates in the GA formalism on the GA computational basis states
$\left\{  1\text{, }i\sigma_{1}\text{, }i\sigma_{2}\text{, }i\sigma
_{3}\right\}  $.\begin{table}[t]
\centering
\begin{tabular}
[c]{c|c|c|c|c|c|c}\hline\hline
\textbf{Single-qubit states} & \textbf{NOT} & \textbf{Phase flip} &
\textbf{Bit and phase flip} & \textbf{Hadamard} & \textbf{Rotation} & $\pi
/8$\textbf{-gate}\\\hline
$1$ & $-i\sigma_{2}$ & $1$ & $i\sigma_{1}$ & $\frac{1-i\sigma_{2}}{\sqrt{2}}$
& $1$ & $1$\\\hline
$i\sigma_{1}$ & $i\sigma_{3}$ & $-i\sigma_{1}$ & $1$ & $\frac{-i\sigma
_{1}+i\sigma_{3}}{\sqrt{2}}$ & $i\sigma_{1}\left(  \cos\theta+i\sigma_{3}%
\sin\theta\right)  $ & $i\sigma_{1}\frac{\left(  1+i\sigma_{3}\right)  }%
{\sqrt{2}}$\\\hline
$i\sigma_{2}$ & $-1$ & $-i\sigma_{2}$ & $i\sigma_{3}$ & $-\frac{1+i\sigma_{2}%
}{\sqrt{2}}$ & $i\sigma_{2}\left(  \cos\theta+i\sigma_{3}\sin\theta\right)  $
& $i\sigma_{2}\frac{\left(  1+i\sigma_{3}\right)  }{\sqrt{2}}$\\\hline
$i\sigma_{3}$ & $i\sigma_{1}$ & $i\sigma_{3}$ & $i\sigma_{2}$ & $\frac
{i\sigma_{1}+i\sigma_{3}}{\sqrt{2}}$ & $i\sigma_{3}$ & $i\sigma_{3}$\\\hline
\end{tabular}
\caption{Geometric algebra description of the action of some of the most
relevant single-qubit quantum gates on the computational basis states
$\left\{  1\text{, }i\sigma_{1}\text{, }i\sigma_{2}\text{, }i\sigma
_{3}\right\}  $.}%
\end{table}\ 

Summing up, in the GA picture of quantum computing, qubits are elements of the
even subalgebra, unitary quantum gates are specified by rotors, and the
bivector $i\sigma_{3}$ controls the usual complex structure of quantum
mechanics. In the GA formalism, quantum gates have a neat geometrical
interpretation. In the ordinary description of quantum gates, a joint
combination of rotations and global phase shifts on the qubit can be employed
to characterize an arbitrary unitary operator on a single qubit as $\hat
{U}=e^{i_{%
\mathbb{C}
}\alpha}R_{\hat{n}}\left(  \theta\right)  $, given some\textbf{\ }%
\emph{real}\textbf{\ }numbers $\alpha$ and $\theta$ along with a\textbf{\ }%
\emph{real} three-dimensional unit vector $\hat{n}=\left(  n_{1}\text{, }%
n_{2}\text{, }n_{3}\right)  $\textbf{.} To illustrate this fact, consider the
Hadamard gate $\hat{H}$ that acts on a single qubit. It satisfies the
relations $\hat{H}\hat{\Sigma}_{1}\hat{H}=\hat{\Sigma}_{3}$ and $\hat{H}%
\hat{\Sigma}_{3}\hat{H}=\hat{\Sigma}_{1}$, with $\hat{H}^{2}=\hat{I}$. Given
these constraints and up to an overall phase, $\hat{H}$ can be viewed as a
$\theta=\pi$ rotation about the axis $\hat{n}=\left(  \hat{n}_{1}+\hat{n}%
_{3}\right)  /\sqrt{2}$ that rotates $\hat{x}$ to $\hat{z}$ and the other way
around. Explicitly, we have $\hat{H}=-i_{%
\mathbb{C}
}R_{\left(  \hat{n}_{1}+\hat{n}_{3}\right)  /\sqrt{2}}\left(  \pi\right)
$\textbf{. }In the GA formalism, rotors are used to handle rotations. The
Hadamard gate, for example, possesses a neat \emph{real}\textbf{\emph{\ }}
geometric interpretation where there is no need for the use of \emph{complex}
numbers. Indeed, it is specified by a rotor $\hat{H}^{\left(
\text{\textrm{GA}}\right)  }=e^{-i\frac{\pi}{2}\frac{\sigma_{1}+\sigma_{3}%
}{\sqrt{2}}}$ that characterizes a rotation by $\pi$ about the $\left(
\sigma_{1}+\sigma_{3}\right)  /\sqrt{2}$ axis.\textbf{\ }It is simple to check
that, up to an overall irrelevant phase shift, the action of the rotor
$\hat{H}^{\left(  \text{GA}\right)  }$ on the $1$\textbf{-}qubit computational
basis states fulfills the transformation rules in Table I. We emphasize that
when a rotor for a rotation by $\pi$ specifies the Hadamard gate, we have
$\hat{H}^{\left(  \text{\textrm{GA}}\right)  2}=-1$. Therefore, it appears
that a reflection rather than a rotation represents more precisely the gate.
When state amplitudes changed by the Hadamard gate are combined with the ones
transformed by different types of gates, the phase difference can become
important. In Ref. \cite{doran1}, it was suggested to treat the Hadamard gate
as a rotation. However, it is now acknowledged the problem with this
viewpoint. Analogously, similar geometric remarks could be developed for the
remaining $1$-qubit gates \cite{doran1}.

\subsection{$2$-Qubit Quantum Computing}

Using the GA formalism, we take into consideration simple circuit models of
quantum computing with $2$-qubit quantum gates. We begin with a simple MSTA
characterization of the maximally entangled $2$-qubit Bell states.

\emph{Geometric Algebra and Bell States. }We display a GA description of the
$2$-qubits Bell states. These states specify a set of four orthonormal
maximally entangled state vectors that represent a basis ($\mathcal{B}%
_{\text{\textrm{Bell}}}$) for the product Hilbert space $%
\mathbb{C}
^{2}\otimes%
\mathbb{C}
^{2}\cong%
\mathbb{C}
^{4}$. Given the $2$-qubit computational basis $\mathcal{B}%
_{\text{computational}}\overset{\text{def}}{=}\left\{  \left\vert
00\right\rangle \text{, }\left\vert 01\right\rangle \text{, }\left\vert
10\right\rangle \text{, }\left\vert 11\right\rangle \right\}  $, the four Bell
states are defined as \cite{NIELSEN},%
\begin{align}
\left\vert 0\right\rangle \otimes\left\vert 0\right\rangle  &  \rightarrow
\left\vert \psi_{\text{\textrm{Bell}}_{1}}\right\rangle \overset{\text{def}%
}{=}\left[  \hat{U}_{\text{CNOT}}\circ\left(  \hat{H}\otimes\hat{I}\right)
\right]  \left(  \left\vert 0\right\rangle \otimes\left\vert 0\right\rangle
\right)  =\frac{1}{\sqrt{2}}\left(  \left\vert 0\right\rangle \otimes
\left\vert 0\right\rangle +\left\vert 1\right\rangle \otimes\left\vert
1\right\rangle \right)  \text{,}\nonumber\\
& \nonumber\\
\left\vert 0\right\rangle \otimes\left\vert 1\right\rangle  &  \rightarrow
\left\vert \psi_{\text{\textrm{Bell}}_{2}}\right\rangle \overset{\text{def}%
}{=}\left[  \hat{U}_{\text{CNOT}}\circ\left(  \hat{H}\otimes\hat{I}\right)
\right]  \left(  \left\vert 0\right\rangle \otimes\left\vert 1\right\rangle
\right)  =\frac{1}{\sqrt{2}}\left(  \left\vert 0\right\rangle \otimes
\left\vert 1\right\rangle +\left\vert 1\right\rangle \otimes\left\vert
0\right\rangle \right)  \text{,}\nonumber\\
& \nonumber\\
\left\vert 1\right\rangle \otimes\left\vert 0\right\rangle  &  \rightarrow
\left\vert \psi_{\text{\textrm{Bell}}_{3}}\right\rangle \overset{\text{def}%
}{=}\left[  \hat{U}_{\text{CNOT}}\circ\left(  \hat{H}\otimes\hat{I}\right)
\right]  \left(  \left\vert 1\right\rangle \otimes\left\vert 0\right\rangle
\right)  =\frac{1}{\sqrt{2}}\left(  \left\vert 0\right\rangle \otimes
\left\vert 0\right\rangle -\left\vert 1\right\rangle \otimes\left\vert
1\right\rangle \right)  \text{,}\nonumber\\
& \nonumber\\
\left\vert 1\right\rangle \otimes\left\vert 1\right\rangle  &  \rightarrow
\left\vert \psi_{\text{\textrm{Bell}}_{4}}\right\rangle \overset{\text{def}%
}{=}\left[  \hat{U}_{\text{CNOT}}\circ\left(  \hat{H}\otimes\hat{I}\right)
\right]  \left(  \left\vert 1\right\rangle \otimes\left\vert 1\right\rangle
\right)  =\frac{1}{\sqrt{2}}\left(  \left\vert 0\right\rangle \otimes
\left\vert 1\right\rangle -\left\vert 1\right\rangle \otimes\left\vert
0\right\rangle \right)  \text{.} \label{so}%
\end{align}
In Eq. (\ref{so}), the operators $\hat{H}$ and $\hat{U}_{\text{CNOT}}$ specify
the Hadamard and the CNOT gates, respectively. The Bell basis $\mathcal{B}%
_{\text{\textrm{Bell}}}$ of $%
\mathbb{C}
^{2}\otimes%
\mathbb{C}
^{2}\cong%
\mathbb{C}
^{4}$ becomes,%
\begin{equation}
\mathcal{B}_{\text{\textrm{Bell}}}\overset{\text{def}}{=}\left\{  \left\vert
\psi_{\text{\textrm{Bell}}_{1}}\right\rangle \text{, }\left\vert
\psi_{\text{\textrm{Bell}}_{2}}\right\rangle \text{, }\left\vert
\psi_{\text{\textrm{Bell}}_{3}}\right\rangle \text{, }\left\vert
\psi_{\text{\textrm{Bell}}_{4}}\right\rangle \right\}  \text{,}%
\end{equation}
where, making use of Eq. (\ref{so}), we have%
\begin{equation}
\left\vert \psi_{\text{\textrm{Bell}}_{1}}\right\rangle =\frac{1}{\sqrt{2}%
}\left(
\begin{array}
[c]{c}%
1\\
0\\
0\\
1
\end{array}
\right)  \text{, }\left\vert \psi_{\text{\textrm{Bell}}_{2}}\right\rangle
=\frac{1}{\sqrt{2}}\left(
\begin{array}
[c]{c}%
0\\
1\\
1\\
0
\end{array}
\right)  \text{, }\left\vert \psi_{\text{\textrm{Bell}}_{3}}\right\rangle
=\frac{1}{\sqrt{2}}\left(
\begin{array}
[c]{c}%
1\\
0\\
0\\
-1
\end{array}
\right)  \text{, }\left\vert \psi_{\text{\textrm{Bell}}_{4}}\right\rangle
=\frac{1}{\sqrt{2}}\left(
\begin{array}
[c]{c}%
0\\
1\\
-1\\
0
\end{array}
\right)  \text{.}%
\end{equation}
Employing Eqs. (\ref{d2}) and (\ref{so}), the Bell states in the GA language
become%
\begin{align}
\left\vert \psi_{\text{\textrm{Bell}}_{1}}\right\rangle  &  \leftrightarrow
\psi_{\text{\textrm{Bell}}_{1}}^{\left(  \text{\textrm{GA}}\right)  }=\frac
{1}{2^{\frac{3}{2}}}\left(  1+i\sigma_{2}^{1}i\sigma_{2}^{2}\right)  \left(
1-i\sigma_{3}^{1}i\sigma_{3}^{2}\right)  \text{, }\left\vert \psi
_{\text{\textrm{Bell}}_{2}}\right\rangle \leftrightarrow\psi
_{\text{\textrm{Bell}}_{2}}^{\left(  \text{\textrm{GA}}\right)  }=-\frac
{1}{2^{\frac{3}{2}}}\left(  i\sigma_{2}^{1}+i\sigma_{2}^{2}\right)  \left(
1-i\sigma_{3}^{1}i\sigma_{3}^{2}\right)  \text{,}\nonumber\\
& \nonumber\\
\left\vert \psi_{\text{\textrm{Bell}}_{3}}\right\rangle  &  \leftrightarrow
\psi_{\text{\textrm{Bell}}_{3}}^{\left(  \text{\textrm{GA}}\right)  }=\frac
{1}{2^{\frac{3}{2}}}\left(  1-i\sigma_{2}^{1}i\sigma_{2}^{2}\right)  \left(
1-i\sigma_{3}^{1}i\sigma_{3}^{2}\right)  \text{, }\left\vert \psi
_{\text{\textrm{Bell}}_{4}}\right\rangle \leftrightarrow\psi
_{\text{\textrm{Bell}}_{4}}^{\left(  \text{\textrm{GA}}\right)  }=\frac
{1}{2^{\frac{3}{2}}}\left(  i\sigma_{2}^{1}-i\sigma_{2}^{2}\right)  \left(
1-i\sigma_{3}^{1}i\sigma_{3}^{2}\right)  \text{.}%
\end{align}
In Fig. $1$, we report a depiction of a quantum circuit for preparing a
maximally entangled $2$-qubit Bell state $\left\vert \psi_{\mathrm{Bell}_{1}%
}\right\rangle $ with a $1$-qubit Hadamard gate and a $2$-qubit
CNOT\ gate.\begin{figure}[t]
\centering
\includegraphics[width=0.35\textwidth] {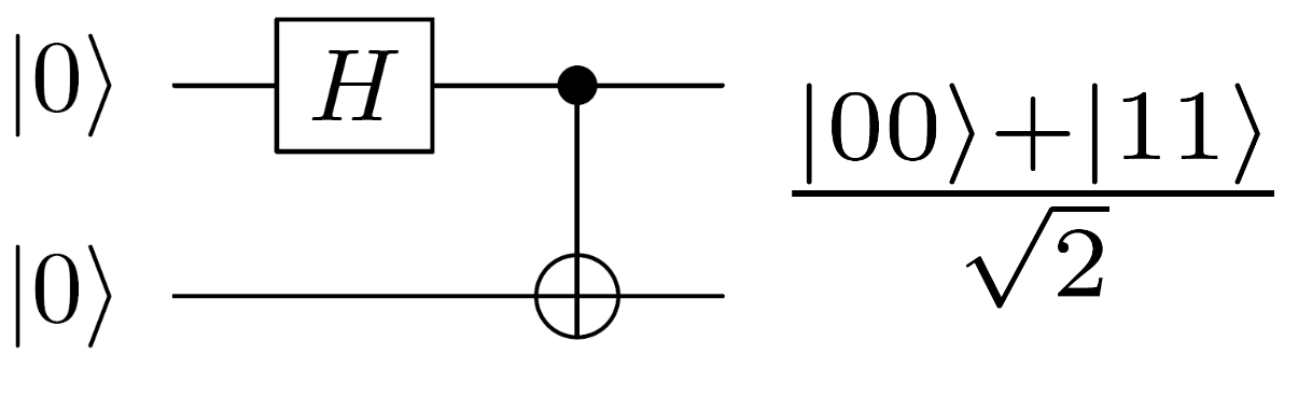}\caption{Schematic depiction of a
quantum circuit for preparing a maximally entangled $2$-qubit Bell state
$\left\vert \psi_{\mathrm{Bell}_{1}}\right\rangle \overset{\text{def}}%
{=}(\left\vert 00\right\rangle +\left\vert 11\right\rangle )/\sqrt{2}$ with a
$1$-qubit Hadamard gate and a $2$-qubit CNOT\ gate. The GA representation of
$\left\vert \psi_{\mathrm{Bell}_{1}}\right\rangle $ is given by $\psi
_{\mathrm{Bell}_{1}}^{\left(  \mathrm{GA}\right)  }\overset{\text{def}}%
{=}\left[  \left(  1+i\sigma_{2}^{1}i\sigma_{2}^{2}\right)  E\right]
/\sqrt{2}$ with $E\overset{\text{def}}{=}\left(  1-i\sigma_{3}^{1}i\sigma
_{3}^{2}\right)  /2$ being the $2$-particle correlator.}%
\end{figure}Interestingly, both abstract spin spaces and abstract index
conventions are unnecessary within the MSTA\ language. Abstract spin spaces
are specified by the complex Hilbert space $\mathcal{H}_{2}^{n}$ of $n$-qubit
quantum states and contain states that must be acted on by quantum unitary
operators. For instance, in the case of Bell states, such operators become the
CNOT gates. Furthermore, the MSTA formalism avoids the use of explicit matrix
representations and, in addition, right or left multiplication by elements
originating from a properly identified geometric algebra play the role of
operators. The proper GA is selected based on the type of qubit quantum states
being acted upon by the operators. This is an additional indication of the
conceptual unification provided by the GA language since \textquotedblleft
spin (qubit) space\textquotedblright\ and \textquotedblleft unitary operators
upon spin space\textquotedblright\ are united, becoming multivectors in real
space. In all honesty, we remark that most GA applications in mathematical
physics exhibit this conceptual unification.

\emph{CNOT Quantum Gate. }Following Ref. \cite{NIELSEN}, the CNOT quantum gate
can be conveniently recast as
\begin{equation}
\hat{U}_{\text{CNOT}}^{12}=\frac{1}{2}\left[  \left(  \hat{I}^{1}+\hat{\Sigma
}_{3}^{1}\right)  \otimes\hat{I}^{2}+\left(  \hat{I}^{1}-\hat{\Sigma}_{3}%
^{1}\right)  \otimes\hat{\Sigma}_{1}^{2}\right]  \text{,} \label{some}%
\end{equation}
with the operator $\hat{U}_{\text{CNOT}}^{12}$ denoting the CNOT gate from
qubit $1$ to qubit $2$. From Eq. (\ref{some}), we have%
\begin{equation}
\hat{U}_{\text{CNOT}}^{12}\left\vert \psi\right\rangle =\frac{1}{2}\left(
\hat{I}^{1}\otimes\hat{I}^{2}+\hat{\Sigma}_{3}^{1}\otimes\hat{I}^{2}+\hat
{I}^{1}\otimes\hat{\Sigma}_{1}^{2}-\hat{\Sigma}_{3}^{1}\otimes\hat{\Sigma}%
_{1}^{2}\right)  \left\vert \psi\right\rangle \text{.} \label{11}%
\end{equation}
Using Eqs. (\ref{impo1}) and (\ref{11}), we get%
\begin{equation}
\hat{I}^{1}\otimes\hat{I}^{2}\left\vert \psi\right\rangle \leftrightarrow
\psi\text{, }\hat{\Sigma}_{3}^{1}\otimes\hat{I}^{2}\left\vert \psi
\right\rangle \leftrightarrow-i\sigma_{3}^{1}\psi J\text{, }\hat{I}^{1}%
\otimes\hat{\Sigma}_{1}^{2}\left\vert \psi\right\rangle \leftrightarrow
-i\sigma_{1}^{2}\psi J\text{, }-\hat{\Sigma}_{3}^{1}\otimes\hat{\Sigma}%
_{1}^{2}\left\vert \psi\right\rangle \leftrightarrow i\sigma_{3}^{1}%
i\sigma_{1}^{2}\psi E\text{.} \label{12}%
\end{equation}
Finally, making use of Eqs. (\ref{11}) and (\ref{12}), the CNOT gate in the GA
language becomes%
\begin{equation}
\hat{U}_{\text{CNOT}}^{12}\left\vert \psi\right\rangle \leftrightarrow\frac
{1}{2}\left(  \psi-i\sigma_{3}^{1}\psi J-i\sigma_{1}^{2}\psi J+i\sigma_{3}%
^{1}i\sigma_{1}^{2}\psi E\right)  \text{.} \label{13}%
\end{equation}

\emph{Controlled-Phase Gate. }From Ref. \cite{NIELSEN}, the action of the
controlled-phase gate $\hat{U}_{\text{CP}}^{12}$ on $\left\vert \psi
\right\rangle \in\mathcal{H}_{2}^{2}$ is,%
\begin{equation}
\hat{U}_{\text{CP}}^{12}\left\vert \psi\right\rangle =\frac{1}{2}\left[
\hat{I}^{1}\otimes\hat{I}^{2}+\hat{\Sigma}_{3}^{1}\otimes\hat{I}^{2}+\hat
{I}^{1}\otimes\hat{\Sigma}_{3}^{2}-\hat{\Sigma}_{3}^{1}\otimes\hat{\Sigma}%
_{3}^{2}\right]  \left\vert \psi\right\rangle \text{.} \label{11a}%
\end{equation}
From Eqs. (\ref{impo1}) and (\ref{11a}), we get%
\begin{equation}
\hat{I}^{1}\otimes\hat{I}^{2}\left\vert \psi\right\rangle \leftrightarrow
\psi\text{, }\hat{\Sigma}_{3}^{1}\otimes\hat{I}^{2}\left\vert \psi
\right\rangle \leftrightarrow-i\sigma_{3}^{1}\psi J\text{, }\hat{I}^{1}%
\otimes\hat{\Sigma}_{3}^{2}\left\vert \psi\right\rangle \leftrightarrow
-i\sigma_{3}^{2}\psi J\text{, }-\hat{\Sigma}_{3}^{1}\otimes\hat{\Sigma}%
_{3}^{2}\left\vert \psi\right\rangle \leftrightarrow i\sigma_{3}^{1}%
i\sigma_{3}^{2}\psi E\text{.} \label{11b}%
\end{equation}
Finally, using Eqs. (\ref{11a}) and (\ref{11b}), the controlled-phase quantum
gate in the GA language reduces to%
\begin{equation}
\hat{U}_{\text{CP}}^{12}\left\vert \psi\right\rangle \leftrightarrow\frac
{1}{2}\left(  \psi-i\sigma_{3}^{1}\psi J-i\sigma_{3}^{2}\psi J+i\sigma_{3}%
^{1}i\sigma_{3}^{2}\psi E\right)  \text{.} \label{14}%
\end{equation}

\emph{SWAP Gate. }From Ref. \cite{NIELSEN}, the action of the SWAP gate
$\hat{U}_{\text{SWAP}}^{12}$ on $\left\vert \psi\right\rangle \in
\mathcal{H}_{2}^{2}$ is,%
\begin{equation}
\hat{U}_{\text{SWAP}}^{12}\left\vert \psi\right\rangle =\frac{1}{2}\left(
\hat{I}^{1}\otimes\hat{I}^{2}+\hat{\Sigma}_{1}^{1}\otimes\hat{\Sigma}_{1}%
^{2}+\hat{\Sigma}_{2}^{1}\otimes\hat{\Sigma}_{2}^{2}+\hat{\Sigma}_{3}%
^{1}\otimes\hat{\Sigma}_{3}^{2}\right)  \left\vert \psi\right\rangle \text{.}
\label{12a}%
\end{equation}
Using Eqs. (\ref{impo1}) and (\ref{12a}), we have%
\begin{equation}
\hat{I}^{1}\otimes\hat{I}^{2}\left\vert \psi\right\rangle \leftrightarrow
\psi\text{, }\hat{\Sigma}_{1}^{1}\otimes\hat{\Sigma}_{1}^{2}\left\vert
\psi\right\rangle \leftrightarrow-i\sigma_{1}^{1}i\sigma_{1}^{2}\psi E\text{,
}\hat{\Sigma}_{2}^{1}\otimes\hat{\Sigma}_{2}^{2}\left\vert \psi\right\rangle
\leftrightarrow-i\sigma_{2}^{1}i\sigma_{2}^{2}\psi E\text{, }\hat{\Sigma}%
_{3}^{1}\otimes\hat{\Sigma}_{3}^{2}\left\vert \psi\right\rangle
\leftrightarrow-i\sigma_{3}^{1}i\sigma_{3}^{2}\psi E\text{.} \label{12b}%
\end{equation}
Finally, employing Eqs. (\ref{12a}) and (\ref{12b}), the SWAP gate in the
GA\ language becomes,%
\begin{equation}
\hat{U}_{\text{SWAP}}^{12}\left\vert \psi\right\rangle \leftrightarrow\frac
{1}{2}\left(  \psi-i\sigma_{1}^{1}i\sigma_{1}^{2}\psi E-i\sigma_{2}^{1}%
i\sigma_{2}^{2}\psi E-i\sigma_{3}^{1}i\sigma_{3}^{2}\psi E\right)  \text{.}
\label{15}%
\end{equation}
In Table II, we display the GA\ description of the action of some of the most
relevant $2$-qubit quantum gates on the GA computational basis $\mathcal{B}%
_{\left[  \mathfrak{cl}^{+}(3)\otimes\mathfrak{cl}^{+}(3)\right]  /E}%
$.\begin{table}[t]
\centering
\begin{tabular}
[c]{c|c|c}\hline\hline
\textbf{Two-qubit gates} & \textbf{Two-qubit states} & \textbf{GA action of
gates on states}\\\hline
CNOT & $\psi$ & $\frac{1}{2}\left(  \psi-i\sigma_{3}^{1}\psi J-i\sigma_{1}%
^{2}\psi J+i\sigma_{3}^{1}i\sigma_{1}^{2}\psi E\right)  $\\\hline
Controlled-Phase Gate & $\psi$ & $\frac{1}{2}\left(  \psi-i\sigma_{3}^{1}\psi
J-i\sigma_{3}^{2}\psi J+i\sigma_{3}^{1}i\sigma_{3}^{2}\psi E\right)  $\\\hline
SWAP & $\psi$ & $\frac{1}{2}\left(  \psi-i\sigma_{1}^{1}i\sigma_{1}^{2}\psi
E-i\sigma_{2}^{1}i\sigma_{2}^{2}\psi E-i\sigma_{3}^{1}i\sigma_{3}^{2}\psi
E\right)  $\\\hline
\end{tabular}
\caption{Geometric algebra description of the action of some of the most
relevant $2$-qubit quantum gates on the GA computational basis $\mathcal{B}%
_{\left[  \mathfrak{cl}^{+}(3)\otimes\mathfrak{cl}^{+}(3)\right]  /E}$.}%
\end{table}Interestingly, two-qubit quantum gates can be geometrically
interpreted by means of rotations. For example, the CNOT gate specifies a
rotation in one qubit space\textbf{\ }conditional\ on the quantum state of a
different qubit it is correlated with. In the GA\ language, this CNOT gate
becomes\textbf{\ }$(\hat{U}_{\text{CNOT}}^{12\left(  \text{\textrm{GA}%
}\right)  })=e^{-i\frac{\pi}{2}\frac{1}{2}\sigma_{1}^{1}\left(  1-\sigma
_{3}^{2}\right)  }$\textbf{. }In particular, this operator acts as a rotation
on the first qubit by an angle $\pi$ about the axis $\sigma_{1}^{1}$ in those
$2$-qubit quantum states in which qubit is located along the $-\sigma_{2}^{3}$
axis. For further technical details on analogous geometrically flavored
considerations for other $2$-qubit gates, we refer to Ref. \cite{somaroo}.

In what follows, we briefly discuss the application of the MSTA formalism to
density matrices for mixed quantum states.

\subsection{Density Operators}

It is known that the statistical aspects of quantum systems can be suitably
described by density matrices and, instead, cannot be specified by means of a
single wave function. For a pure state $\left\vert \psi\right\rangle
=\alpha\left\vert 0\right\rangle +\beta\left\vert 1\right\rangle $, the
density matrix can be recast as%
\begin{equation}
\hat{\rho}_{\text{pure}}=\left\vert \psi\right\rangle \left\langle
\psi\right\vert =\left(
\begin{array}
[c]{cc}%
\alpha\alpha^{\ast} & \alpha\beta^{\ast}\\
\beta\alpha^{\ast} & \beta\beta^{\ast}%
\end{array}
\right)  \text{.}%
\end{equation}
Importantly, the expectation value $\left\langle \hat{O}\right\rangle $ of any
observable $\hat{O}$ with respect to a given normalized quantum state
$\left\vert \psi\right\rangle $ can be derived from $\hat{\rho}_{\text{pure}}$
by noting that $\left\langle \hat{O}\right\rangle =\left\langle \psi|\hat
{O}|\psi\right\rangle =$\textrm{tr}$(\hat{\rho}_{\text{pure}}\hat{O})$. The
formulation of $\hat{\rho}_{\text{pure}}$ in the GA\ language is given by%
\begin{equation}
\hat{\rho}_{\text{pure}}\rightarrow\rho_{\text{pure}}^{\left(
\text{\textrm{GA}}\right)  }=\psi\frac{1}{2}\left(  1+\sigma_{3}\right)
\psi^{\dagger}=\frac{1}{2}\left(  1+s\right)  \text{,} \label{nick}%
\end{equation}
with $s$ denoting the spin vector defined as $s\overset{\text{def}}{=}%
\psi\sigma_{3}\psi^{\dagger}$ \cite{doran96}. From Eq. (\ref{nick}), we
observe that $\rho_{\text{pure}}^{\left(  \text{\textrm{GA}}\right)  }$ is
simply the sum of a scalar and a vector from a geometric standpoint. In
standard quantum mechanics, a density matrix for a mixed quantum state
$\hat{\rho}_{\text{mixed}}$ can be expressed in terms of a weighted sum of the
density matrices for the pure quantum states as%
\begin{equation}
\hat{\rho}_{\text{mixed}}=\sum_{j=1}^{n}\hat{\rho}_{j}=\sum_{j=1}^{n}%
p_{j}\left\vert \psi_{j}\right\rangle \left\langle \psi_{j}\right\vert
\text{,} \label{rox}%
\end{equation}
where $\left\{  p_{j}\right\}  _{j=1,...,n}$ \ is a set of (real)
probabilities normalized to one (i.e., $p_{1}+$...$+p_{n}=1$\textbf{,
}where\textbf{ }$0\leq p_{j}\leq1$ for any $j\in\left\{  1,...,n\right\}  $).
In the GA\ language, given that the addition operation is well-defined,
$\hat{\rho}_{\text{mixed}}$ in Eq. (\ref{rox}) can be expressed in terms of a
sum as%
\begin{equation}
\hat{\rho}_{\text{mixed}}\rightarrow\rho_{\text{mixed}}^{\left(
\text{\textrm{GA}}\right)  }=\frac{1}{2}\sum_{j=1}^{n}\left(  p_{j}+p_{j}%
s_{j}\right)  =\frac{1}{2}\left(  1+P\right)  \text{.} \label{love}%
\end{equation}
The quantity $P$ in Eq. (\ref{love}) denotes the average spin vector (i.e.,
the ensemble-average polarization vector) with magnitude $\left\Vert
P\right\Vert $ satisfying the inequality $\left\Vert P\right\Vert \leq1$. The
magnitude $\left\Vert P\right\Vert $ is a measure of the degree of alignment
among the unit polarization vectors $\left\{  s_{j}\right\}  $ of the
individual elements of the ensemble. For correctness, we emphasize that
$\rho_{\text{mixed}}^{\left(  \text{\textrm{GA}}\right)  }$ in Eq.
(\ref{love}) is the GA\ description of a density operator for an ensemble of
identical and non-interacting quantum bits. In general, one could take into
consideration expressions of density operators for multi-qubit systems
characterized by interacting quantum bits. In the MSTA formalism, the density
matrix of $n$-interacting qubits can be recast as%
\begin{equation}
\rho_{\text{multi-qubit}}^{\left(  \text{\textrm{GA}}\right)  }=\overline
{\left(  \psi E_{n}\right)  E_{+}\left(  \psi E_{n}\right)  ^{\sim}}\text{.}
\label{love1}%
\end{equation}
In Eq. (\ref{love1}), $E_{n}$ denotes the $n$-particle correlator, while
$E_{+}\overset{\text{def}}{=}E_{+}^{1}E_{+}^{2}$...$E_{+}^{n}$ describes the
geometric product of $n$-idempotents with $E_{\pm}^{k}\overset{\text{def}}%
{=}(1\pm\sigma_{3}^{k})/2$ and $k=1$,..., $n$. Finally, while the tilde symbol
\textquotedblleft$\sim$\textquotedblright\ is used to describe the space-time
reverse, the over-line in Eq. (\ref{love1})\ signifies the ensemble-average.
For further technicalities on the GA approach to density matrices for general
quantum systems, we suggest Ref. \cite{doran1}.

\section{Universality of quantum gates with geometric algebra}

Employing the results obtained in Section III and, in addition, formulating a
GA perspective on the Lie algebras $\mathrm{SO}\left(  3\right)  $ and
$\mathrm{SU}\left(  2\right)  $ that relies on the rotor group $\mathrm{Spin}%
^{+}\left(  3\text{, }0\right)  $ formalism, we discuss in this section a
GA-based version of the universality of quantum gates proof as originally
proposed by Boykin and collaborators in Refs. \cite{B99,B00}. We begin with
the introduction of the rotor group $\mathrm{Spin}^{+}\left(  3\text{,
}0\right)  $. We then bring in some universal sets of quantum gates. Finally,
we discuss our GA-based proof of universality in quantum computing.

\subsection{$\mathrm{SO}(3)$,\textrm{ }$\mathrm{SU}(2)$\textrm{, and
}$\mathrm{Spin}^{+}(3$, $0)$}

Motivated by the fact that the proofs in Refs. \cite{B99,B00} depend in a
significant manner on rotations in three-dimensional space and, in addition,
on the local isomorphism between $\mathrm{SO}(3)$ and $\mathrm{SU}(2)$, we
briefly show how these Lie groups can be described in the GA\ language in
terms of the rotor group $\mathrm{Spin}^{+}\left(  3\text{, }0\right)  $.

\subsubsection{Preliminaries on $\mathrm{SO}(3)$ and $\mathrm{SU}(2)$}

Three-dimensional Lie groups are very important in physics \cite{morton}. In
this regard, the three-dimensional Lie groups $\mathrm{SO}(3)$ and
$\mathrm{SU}(2)$ with Lie algebras $\mathfrak{so(}3\mathfrak{)}$ and
$\mathfrak{su(}2\mathfrak{)}$, respectively, are two physically significant
groups. The group $\mathrm{SO}(3)$ denotes the group of orthogonal
transformations with determinant equal to one (i.e., rotations of
three-dimensional space) and is defined by,%
\begin{equation}
\mathrm{SO}(3)\overset{\text{def}}{=}\left\{  M\in GL(3\text{, }%
\mathbb{R}
):MM^{t}=M^{t}M=I_{3\times3}\text{, }\det M=1\right\}  \text{.} \label{sob}%
\end{equation}
In Eq. (\ref{sob}), $GL(3$, $%
\mathbb{R}
)$ is the general linear group specified by the set of non-singular linear
transformations in $%
\mathbb{R}
^{3}$ characterized by $3\times3$ non singular matrices with real entries. The
letter \textquotedblleft$t$\textquotedblright, instead, means the transpose of
a matrix. The group $\mathrm{SU}(2)$ is the special unitary group all
$2\times2$ unitary complex matrices with determinant equal to one. It is
defined as,%
\begin{equation}
\mathrm{SU}(2)\overset{\text{def}}{=}\left\{  M\in GL(2\text{, }%
\mathbb{C}
):MM^{\dagger}=M^{\dagger}M=I_{2\times2}\text{, }\det M=1\right\}  \text{.}
\label{sob1}%
\end{equation}
In Eq. (\ref{sob1}), $GL(2$, $%
\mathbb{C}
)$ denotes the set of non-singular linear transformations in $%
\mathbb{C}
^{2}$ specified by $2\times2$ non singular matrices with complex entries,
while the symbol \textquotedblleft$\dagger$\textquotedblright\ signifies the
Hermitian conjugation operation. Interestingly, while the Lie algebras
$\mathfrak{so(}3\mathfrak{)}$ and $\mathfrak{su(}2\mathfrak{)}$ are isomorphic
(i.e., $\mathfrak{so(}3\mathfrak{)}\cong\mathfrak{su(}2\mathfrak{)}$), the Lie
groups $\mathrm{SO}(3)$ and $\mathrm{SU}(2)$ are only \emph{locally}
isomorphic. This means that they differ at a global level (i.e., far from
identity), despite the fact that they are not distinguishable in terms of
infinitesimal transformations. This distinguishability at the global level
implies that the $\mathrm{SO}(3)$ and $\mathrm{SU}(2)$ do no give rise to a
pair of isomorphic groups. In particular, this distinguishability manifests
itself in the fact that while a rotation by $2\pi$ is the same as the identity
in $\mathrm{SO}(3)$, the $\mathrm{SU}(2)$ group is periodic exclusively under
rotations by $4\pi$. This implies that while it is an unacceptable
representation of $\mathrm{SO}(3)$, a quantity that acquires a minus sign
under the action of a rotation by an angle equal to $2\pi$ represents an
acceptable representation of $\mathrm{SU}(2)$. Interestingly, as pointed out
in Ref. \cite{maggiore}, spin-$1/2$ particles (or, alternatively, qubits) need
to be rotated by $720^{0}$ (i.e., $4\pi$ radians) to return to the original
state. Moreover, while $\mathrm{SU}(2)$ is topologically equivalent to the
$3$-sphere $\mathcal{S}^{3}$, $\mathrm{SO}(3)$ is topologically equivalent to
the projective space $%
\mathbb{R}
P^{3}$. For completeness, note that $%
\mathbb{R}
P^{3}$ originates from $\mathcal{S}^{3}$ once one identifies pairs of
antipodal points. These comparative remarks between the groups $\mathrm{SO}%
(3)$ and $\mathrm{SU}(2)$ imply that the groups that are actually isomorphic
are the quotient group $\mathrm{SU}(2)/%
\mathbb{Z}
_{2}$ and $\mathrm{SO}(3)$ (i.e., $\mathrm{SU}(2)/%
\mathbb{Z}
_{2}\cong\mathrm{SO}(3)$). From a formal mathematical standpoint, there exists
an unfaithful representation $\varkappa$ of $\mathrm{SU}(2)$ as a group of
rotations in $%
\mathbb{R}
^{3}$,%
\begin{equation}
\varkappa:\mathrm{SU}(2)\ni U_{\mathrm{SU}(2)}(\vec{A}\text{, }\theta
)\overset{\text{def}}{=}\exp(\frac{\vec{\Sigma}}{2i_{%
\mathbb{C}
}}\cdot\vec{A}\theta)\mapsto R_{\mathrm{SO}(3)}(\vec{A}\text{, }%
\theta)\overset{\text{def}}{=}\exp(\vec{E}\cdot\vec{A}\theta)\in
\mathrm{SO}(3)\text{,} \label{70}%
\end{equation}
for any vector $\vec{A}=\left(  A_{1}\text{, }A_{2}\text{, }A_{3}\right)  $ in
$%
\mathbb{R}
^{3}$. For mathematical accuracy, we emphasize that the employment of the
dot-notation in Eq. (\ref{70})\ (and, in addition, in the following Eqs.
(\ref{72}), (\ref{83}), (\ref{jj})) represents an abuse of notation for the
Euclidean inner product. As a matter of fact, while $\vec{A}$\ is simply a
vector in $%
\mathbb{R}
^{3}$, the quantity $\vec{\Sigma}$ \ specifies the vector of Pauli operators
that act on a two-dimensional complex Hilbert space.\textbf{\ }Note that the
vector $\vec{E}=\left(  E_{1}\text{, }E_{2}\text{, }E_{3}\right)  $ in Eq.
(\ref{70}) determines a basis of infinitesimal generators of the Lie algebra
$\mathfrak{so(}3\mathfrak{)}$ of the group $\mathrm{SO}(3)$,%
\begin{equation}
E_{1}\overset{\text{def}}{=}\left(
\begin{array}
[c]{ccc}%
0 & 0 & 0\\
0 & 0 & -1\\
0 & 1 & 0
\end{array}
\right)  \text{, }E_{2}\overset{\text{def}}{=}\left(
\begin{array}
[c]{ccc}%
0 & 0 & 1\\
0 & 0 & 0\\
-1 & 0 & 0
\end{array}
\right)  \text{, }E_{3}\overset{\text{def}}{=}\left(
\begin{array}
[c]{ccc}%
0 & -1 & 0\\
1 & 0 & 0\\
0 & 0 & 0
\end{array}
\right)  \text{.} \label{mado}%
\end{equation}
The matrices $\left\{  E_{j}\right\}  $ with $j\in\left\{  1,2,3\right\}  $ in
Eq. (\ref{mado}) fulfill the commutation relations, $\left[  E_{l}\text{,
}E_{m}\right]  =\varepsilon_{lmk}E_{k}$ with $\varepsilon_{lmk}$ being the
usual Levi-Civita symbol. Alternatively, the infinitesimal generators of the
Lie algebra $\mathfrak{su(}2\mathfrak{)}$ of the special unitary group
$\mathrm{SU}(2)$ are determined by $i_{%
\mathbb{C}
}\vec{\Sigma}=\left(  i_{%
\mathbb{C}
}\Sigma_{1}\text{, }i_{%
\mathbb{C}
}\Sigma_{2}\text{, }i_{%
\mathbb{C}
}\Sigma_{3}\right)  $. These generators satisfy the commutation relations
given by $\left[  \Sigma_{l}\text{, }\Sigma_{m}\right]  =2i_{%
\mathbb{C}
}\varepsilon_{lmk}\Sigma_{k}$. It is worthwhile mentioning that these latter
commutation relations are identical to those for $\mathrm{SO}(3)$ once one
employs $\Sigma_{l}/2i_{%
\mathbb{C}
}$ as a new basis for the algebra $\mathfrak{su(}2\mathfrak{)}$. The map
$\varkappa$ in Eq. (\ref{70}) is exactly two-to-one. Therefore, \ to a
rotation of $%
\mathbb{R}
^{3}$ about an axis specified by a unit vector $\vec{A}$ through an angle of
$\theta$ radians, there correspond two $2\times2$ unitary matrices with
determinant equal to one,%
\begin{equation}
\exp(\frac{\vec{\Sigma}}{2i_{%
\mathbb{C}
}}\cdot\vec{A}\theta)\text{ and, }\exp[\frac{\vec{\Sigma}}{2i_{%
\mathbb{C}
}}\cdot\vec{A}\left(  \theta+2\pi\right)  ]\text{.} \label{72}%
\end{equation}
Put differently, not only $\mathrm{SO}(3)$ possesses the traditional
representation in terms of $3\times3$ matrices, it also enjoys a double-valued
representation by means of $2\times2$ matrices acting on $%
\mathbb{C}
^{2}$. In this respect, spinors can be simply viewed as the complex vectors $(%
\begin{array}
[c]{cc}%
\psi^{1} & \psi^{2}%
\end{array}
)^{t}\in%
\mathbb{C}
^{2}$ on which $\mathrm{SO}(3)$ operates in this double-valued manner. From a
mathematical standpoint, $\mathrm{SU}(2)$ offers in a natural manner a spinor
representation of the $2$-fold cover of the group $\mathrm{SO}(3)$. Note that
$\mathrm{SU}(2)$ is known as the the spin group $\mathrm{Spin}(3)$ when it is
viewed as the $2$-fold cover of $\mathrm{SO}(3)$. Representing
three-dimensional rotations by means of two-dimensional unitary
transformations is extraordinarily effective. In the framework of quantum
computing, this is particularly correct when demonstrating particular circuit
identities, when characterizing arbitrary $1$-qubit states and, finally, in
constructing the Hardy state \cite{mermin}. To say it all, this representation
has a remarkable role in the proof of universality of quantum gates as
proposed by Boykin and collaborators in Refs. \cite{B99,B00}. In particular,
the surjective homeomorphism $\varkappa$ in Eq. (\ref{70}) represents a
formidable instrument for studying the product of two or more rotations. This
is justified by the fact that Pauli matrices fulfill uncomplicated product
rules, $\Sigma_{l}\Sigma_{m}=\delta_{lm}+i_{%
\mathbb{C}
}\varepsilon_{lmk}\Sigma_{k}$. Unfortunately, the infinitesimal generators
$\left\{  E_{l}\right\}  $ with $l\in\left\{  1\text{, }2\text{, }3\right\}  $
of $\mathfrak{so(}3\mathfrak{)}$ do not satisfy such simple product relations
and, for instance, $E_{1}^{2}=\mathrm{diag}\left(  0\text{, }-1\text{,
}-1\right)  $.

Having discussed some links between $\mathrm{SO}(3)$ and $\mathrm{SU}(2)$, we
are now ready to introduce the group $\mathrm{Spin}^{+}(3$, $0)$.

\subsubsection{Preliminaries on $\mathrm{Spin}^{+}(3$, $0)$}

In the GA\ language, rotations are described by means of \emph{rotors }and
they represent one of the most significant applications of geometric algebra.
Moreover, Lie groups and Lie algebras can be conveniently studied in terms of
rotors. In the following, we present some definitions. For further details on
the Clifford algebras, we refer to Ref. \cite{porto}.

Assume that $\mathcal{G}\left(  p\text{, }q\right)  $ specifies the GA of a
space with signature $\left(  p\text{, }q\right)  $, where $p+q=n$ and $n$ is
the dimensionality of the space. Assume, in addition, that $\mathcal{V}$ is
the space whose elements are grade-$1$ multivectors. Then, \textrm{
}$\mathrm{Pin}\left(  p\text{, }q\right)  $ is the so-called pin group with
respect to the geometric product and is given by,%
\begin{equation}
\mathrm{Pin}\left(  p\text{, }q\right)  \overset{\text{def}}{=}\left\{
M\in\mathcal{G}\left(  p\text{, }q\right)  :MaM^{-1}\in\mathcal{V}\text{
}\forall a\in\mathcal{V}\text{, }MM^{\dagger}=\pm1\right\}  \text{,}%
\end{equation}
with \textquotedblleft$\dagger$\textquotedblright\ specifying the GA reversion
operation where, for example, $\left(  a_{1}a_{2}\right)  ^{\dagger}%
=a_{2}a_{1}$. The elements of the pin group $\mathrm{Pin}\left(  p\text{,
}q\right)  $ can be partitioned into odd-grade and even-grade multivectors.
The even-grade elements $\left\{  S\right\}  $ of the pin group $\mathrm{Pin}%
\left(  p\text{, }q\right)  $ generates a subgroup known as the spin group
$\mathrm{Spin}\left(  p\text{, }q\right)  $,%
\begin{equation}
\mathrm{Spin}\left(  p\text{, }q\right)  \overset{\text{def}}{=}\left\{
S\in\mathcal{G}_{+}\left(  p\text{, }q\right)  :SaS^{-1}\in\mathcal{V}\text{
}\forall a\in\mathcal{V}\text{, }SS^{\dagger}=\pm1\right\}  \text{,}%
\end{equation}
with $\mathcal{G}_{+}\left(  p\text{, }q\right)  $ being the even subalgebra
of $\mathcal{G}\left(  p\text{, }q\right)  $. Then, rotors are nothing but
multivectors $\left\{  R\right\}  $ of the spin group $\mathrm{Spin}\left(
p\text{, }q\right)  $ that fulfill the additional constraining condition
$RR^{\dagger}=+1$. These elements specify the so-called rotor group
$\mathrm{Spin}^{+}\left(  p\text{, }q\right)  $ defined as,%
\begin{equation}
\mathrm{Spin}^{+}\left(  p\text{, }q\right)  \overset{\text{def}}{=}\left\{
R\in\mathcal{G}_{+}\left(  p\text{, }q\right)  :RaR^{\dagger}\in
\mathcal{V}\text{ }\forall a\in\mathcal{V}\text{, }RR^{\dagger}=+1\right\}
\text{.}%
\end{equation}
For spaces like the Euclidean spaces, $\mathrm{Spin}\left(  n\text{,
}0\right)  =\mathrm{Spin}^{+}\left(  n\text{, }0\right)  $. For such
scenarios, the spin group $\mathrm{Spin}\left(  p\text{, }q\right)  $ and the
rotor group $\mathrm{Spin}^{+}\left(  p\text{, }q\right)  $ cannot be distinguished.

In the GA language, the double-sided half-angle transformation law that
specifies the rotation of a vector $a$ by an angle $\theta$ in the plane
spanned by two unit vectors $m$ and $n$ is given by
\begin{equation}
a\rightarrow a^{\prime}\overset{\text{def}}{=}RaR^{\dagger}\text{.}
\label{rotorlaw}%
\end{equation}
The rotor $R$ in Eq. (\ref{rotorlaw}) is given by,%
\begin{equation}
R\overset{\text{def}}{=}nm=n\cdot m+n\wedge m=\exp(-B\frac{\theta}{2})\text{,}
\label{bive}%
\end{equation}
with the bivector $B$ in Eq. (\ref{bive}) being such that,%
\begin{equation}
B\overset{\text{def}}{=}\frac{m\wedge n}{\sin(\frac{\theta}{2})}\text{ and,
}B^{2}=-1\text{.}%
\end{equation}
Thanks to the existence of the geometric product in the GA setting, rotors
offer a unique ways of characterizing rotations in geometric algebra. From Eq.
(\ref{bive}), observe that rotors are mixed-grade multivectors since they are
specified by the geometric product of two unit vectors. Since no special
significance can be assigned to the separate scalar and bivector terms, the
rotor has no meaning on its own. However, observe that the exponential of a
bivector always returns to a rotor and, in addition, all rotors near the
origin can be recast in terms of the exponential of a bivector. Therefore,
since the bivector $B$ has a clear geometric meaning, when the rotor $R$ is
expressed in terms of the exponential of the bivector $B$, both $R$ and the
vector $RaR^{\dagger}$ gain a clear geometrically neat significance. Once
again, this mathematical picture provides an additional illustrative example
of one of the hallmarks of GA. Specifically, both geometrically meaningful
objects (vectors and planes, for instance) and the elements (operators, for
instance) that operate on them (in this example, rotors $\left\{  R\right\}  $
or bivectors $\left\{  B\right\}  $) belong to the same geometric Clifford
algebra. Observe that there is a two-to-one mapping between rotors and
rotations, since $R$ and $-R$ lead to the same rotation. In Fig. $2$, inspired
by the graphical depictions by Doran and Lasenby in Ref. \cite{dl}, we
illustrate a rotation in three dimensions from a geometric algebra
viewpoint.\textbf{ }We stress that one usually thinks of rotations as taking
place around an axis in three-dimensions, a concept that does not generalize
straightforwardly to any dimension. However, the GA language leads us to
regard rotations as taking place in a plane embedded in a higher dimensional
space. Therefore, rotations are described by equations that are valid in
arbitrary dimensions.\textbf{ }\begin{figure}[t]
\centering
\includegraphics[width=0.75\textwidth] {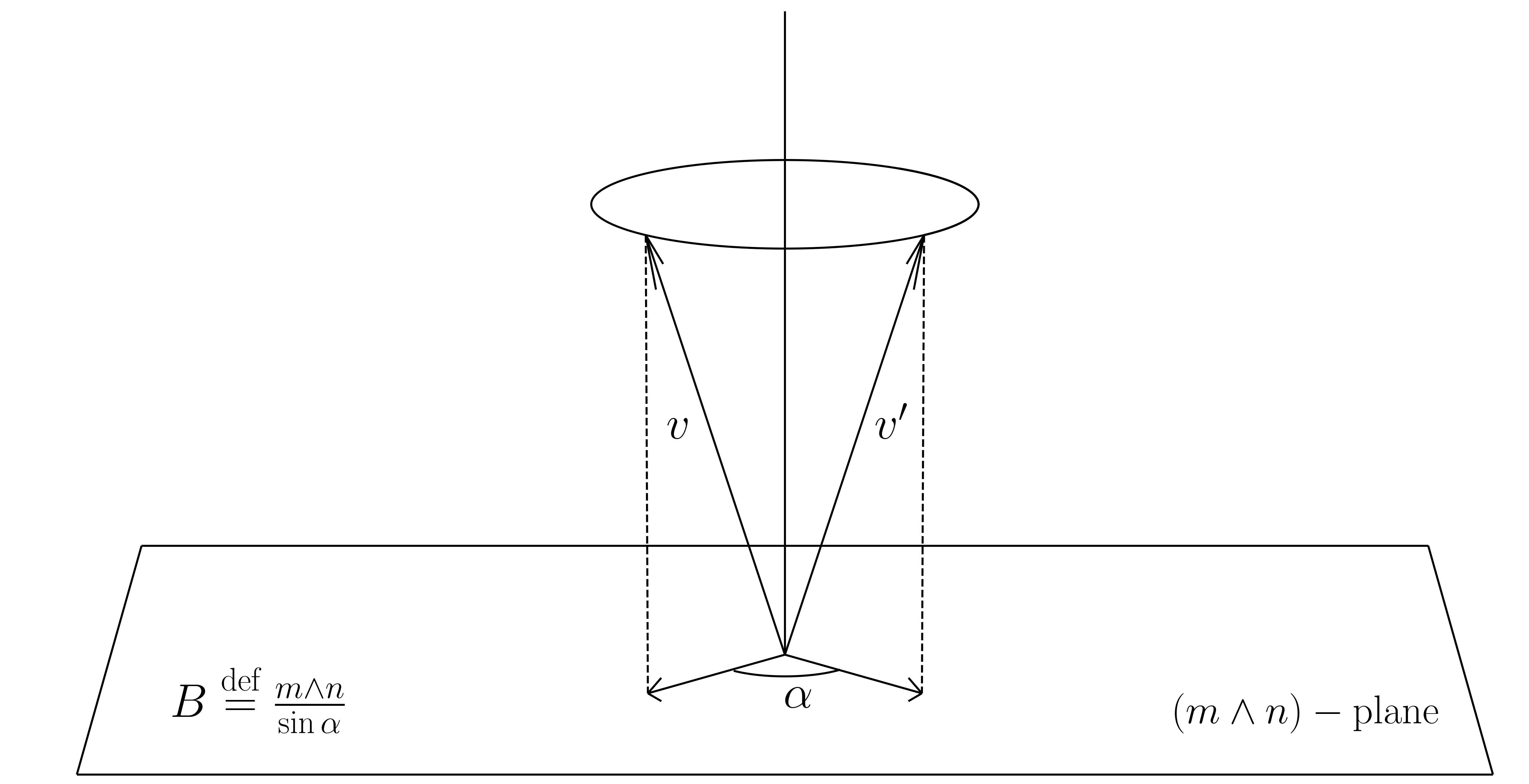}\caption{Schematic depiction of a
rotation in three dimensions from a geometric algebra viewpoint. The vector
$v$ is rotated through an angle $\alpha$ in the $m\wedge n$ plane with $m\cdot
n=\cos\left(  \alpha\right)  $ specified by a unit bivector $B\overset
{\text{def}}{=}\left(  m\wedge n\right)  /\sin(\alpha)$ such that $B^{2}=-1$.
After the rotation, $v$ becomes $v^{\prime}=RvR^{\dagger}$ with $R\overset
{\text{def}}{=}\exp(-B\alpha/2)$ being the rotor describing the rotation in
terms of the $m\wedge n$ plane and the rotation angle $\alpha$.}%
\end{figure}From a formal mathematical standpoint, the rotor group
$\mathrm{Spin}^{+}\left(  p\text{, }q\right)  $ furnishes a double-cover
representation of the rotation group $\mathrm{SO}(n)$. The Lie algebra of the
rotor group $\mathrm{Spin}^{+}\left(  3\text{, }0\right)  $ is determined by
means of the bivector algebra relations,%
\begin{equation}
\left[  B_{l}\text{, }B_{m}\right]  =2B_{l}\times B_{m}=-2\varepsilon
_{lmk}B_{k}\text{,}%
\end{equation}
with \textquotedblleft$\times$\textquotedblright\ denoting the commutator
product between two multivectors in GA framework. Moreover, the bivectors
$\left\{  B_{j}\right\}  $ with $j\in\left\{  1\text{, }2\text{, }3\right\}  $
are defined as
\begin{equation}
B_{1}\overset{\text{def}}{=}\sigma_{2}\sigma_{3}=i\sigma_{1}\text{, }%
B_{2}\overset{\text{def}}{=}\sigma_{3}\sigma_{1}=i\sigma_{2}\text{, }%
B_{3}\overset{\text{def}}{=}\sigma_{1}\sigma_{2}=i\sigma_{3}\text{.}%
\end{equation}
Note that the space of bivectors is closed under the commutator product
\textquotedblleft$\times$\textquotedblright, given the fact that the
commutator of a first bivector with a second bivector produces a third
bivector. This closed algebra, in turn, specifies the Lie algebra of the
corresponding rotor group $\mathrm{Spin}^{+}\left(  p\text{, }q\right)  $. The
act of exponentiation generates the group structure (see Eq. (\ref{bive})).
Moreover, note that the product of bivectors fulfills the following relations,%
\begin{equation}
B_{l}B_{m}=-\delta_{lm}-\varepsilon_{lmk}B_{k}\text{.} \label{yoyo}%
\end{equation}
The antisymmetric part of $B_{l}B_{m}$ in Eq. (\ref{yoyo}) is a bivector,
whereas the symmetric part of this product denotes a scalar quantity. As a
concluding remark, we emphasize that the algebra of the generators of the
quaternions is like the algebra of bivectors in GA. For this reason, bivectors
correspond to quaternions in the GA language. In Table III, we report in a
schematic fashion a comparative description of $\mathrm{SO}(3)$,
$\mathrm{SU}(2)$, and $\mathrm{Spin}^{+}\left(  p\text{, }q\right)
$.\begin{table}[t]
\centering
\begin{tabular}
[c]{c|c|c|c}\hline\hline
\textbf{Lie groups} & \textbf{Lie algebras} & \textbf{Product rules} &
\textbf{Operator,\ Vectors}\\\hline
$\mathrm{SO}(3)$ & $\left[  E_{l}\text{, }E_{m}\right]  =\varepsilon
_{lmk}E_{k}$ & Not useful & Orthogonal transformations, vectors in $\mathbb{R}
^{3}$\\\hline
$\mathrm{SU}(2)$ & $\left[  \Sigma_{l}\text{, }\Sigma_{m}\right]
=2i_{\mathbb{C} }\varepsilon_{lmk}\Sigma_{k}$ & $\Sigma_{l}\Sigma_{m}%
=\delta_{lm}+i_{\mathbb{C} }\varepsilon_{lmk}\Sigma_{k}$ & Unitary operators,
spinors\\\hline
$\mathrm{Spin}^{+}\left(  3\text{, }0\right)  $ & $\left[  B_{l}\text{, }%
B_{m}\right]  =-2\varepsilon_{lmk}B_{k}$ & $B_{l}B_{m}=-\delta_{lm}%
-\varepsilon_{lmk}B_{k}$ & Rotors (or, bivectors), multivectors\\\hline
\end{tabular}
\caption{Schematic description of the relevant relations among $\mathrm{SO}%
(3)$, $\mathrm{SU}(2)$ and, the rotor group $\mathrm{Spin}^{+}\left(  3\text{,
}0\right)  $.}%
\end{table}Summing up, two main aspects of the GA language become visible.
First, unlike when struggling with matrices, GA offers a very neat and
powerful technique to describe rotations. Second, both geometrically
significant quantities (vectors and planes, for example) and the elements
(i.e., operators) that act on them (in our discussion, rotors $\left\{
R\right\}  $ or bivectors $\left\{  B\right\}  $) are members of the very same
GA.\textbf{ }This second feature is a consequence of the fact that one of the
main practical goals of the GA approach is to carry out calculations without
ever needing to employ an explicit matrix representation. Focusing for
simplicity on the quantum theory of spin-$1/2$ particles, operators are
objects in quantum isospace that act on two-component complex spinors that
belong to two-dimensional complex vector spaces. In typical quantum-mechanical
calculations, one fixes a basis of this Hilbert space to find a matrix
representation of the operator acting on an arbitrary quantum state vector
expressed as a linear combination of the basis vectors. Instead, the GA
description of the quantum mechanics of qubits is coordinate-free and all
operations involving spinors occur without abandoning the GA\ of space (i.e.,
the Pauli algebra).

Having introduced the group\textbf{ }$\mathrm{Spin}^{+}(3$\textbf{, }%
$0)$\textbf{, }we can discuss the concept of universal quantum gate.

\subsection{Universal Quantum Gates}

Spins are discrete quantum variables that can represent both inputs and
outputs of suitable input-output devices such as quantum computational gates.
Indeed, recollect that a finite rotation can be used to express an arbitrary
$2\times2$ unitary matrix with determinant equal to one,%
\begin{equation}
\hat{U}_{\mathrm{SU}\left(  2\right)  }\left(  \hat{n}\text{, }\theta\right)
\overset{\text{def}}{=}e^{-i_{%
\mathbb{C}
}\frac{\theta}{2}\hat{n}\cdot\vec{\Sigma}}=\hat{I}\cos(\frac{\theta}{2})-i_{%
\mathbb{C}
}\hat{n}\cdot\vec{\Sigma}\sin(\frac{\theta}{2})\text{.} \label{83}%
\end{equation}
For this reason, we are allowed to view a qubit as the state of a spin-$1/2$
particle. In addition, we can regard an arbitrary quantum gate, expressed as a
unitary transformation that acts on the state, as a rotation of the spin
(modulo an overall phase factor). When any quantum computational task can be
accomplished with arbitrary precision thanks to networks that consist solely
of replicas of gates from that set, such a set of gates is known to be
\emph{adequate}. In the case in which a network characterized by replicas of
only one gate can be used to perform any quantum computation, such a gate
expresses an adequate set and, in particular, is known to be \emph{universal}.
The Deutsch three-bit gate is an example of a universal quantum gate
\cite{D89}. This three-bit gate has a $8\times8$ unitary matrix representation
specified by a matrix $\mathcal{D}_{\text{\textrm{universal}}}^{\left(
\text{\textrm{Deutsch}}\right)  }\left(  \gamma\right)  $. With respect to the
network's computational basis $\mathcal{B}_{\mathcal{H}_{2}^{3}}%
\overset{\text{def}}{=}\left\{  \left\vert 000\right\rangle \text{,}\left\vert
100\right\rangle \text{, }\left\vert 010\right\rangle \text{, }\left\vert
001\right\rangle \text{, }\left\vert 110\right\rangle \text{, }\left\vert
101\right\rangle \text{, }\left\vert 011\right\rangle \text{, }\left\vert
111\right\rangle \right\}  $, $\mathcal{D}_{\text{\textrm{universal}}%
}^{\left(  \text{\textrm{Deutsch}}\right)  }\left(  \gamma\right)  $ becomes%
\begin{equation}
\mathcal{D}_{\text{\textrm{universal}}}^{\left(  \text{\textrm{Deutsch}%
}\right)  }\left(  \gamma\right)  \overset{\text{def}}{=}\left(
\begin{array}
[c]{cc}%
I_{6\times6} & O_{6\times2}\\
O_{2\times6} & D_{2\times2}\left(  \gamma\right)
\end{array}
\right)  \text{,} \label{dg1}%
\end{equation}
with $I_{l\times l}$ being the $l\times l$ identity matrix, $O_{m\times k}$
denoting the $m\times k$ null matrix, and the matrix $D_{2\times2}\left(
\gamma\right)  $ being given by%
\begin{equation}
D_{2\times2}\left(  \gamma\right)  \overset{\text{def}}{=}\left(
\begin{array}
[c]{cc}%
i_{%
\mathbb{C}
}\cos(\frac{\pi\gamma}{2}) & \sin(\frac{\pi\gamma}{2})\\
\sin(\frac{\pi\gamma}{2}) & i_{%
\mathbb{C}
}\cos(\frac{\pi\gamma}{2})
\end{array}
\right)  \text{.}%
\end{equation}
From Eq. (\ref{dg1}), we note that the Deutsch gate is determined by the
parameter $\gamma$ which can assume any irrational value. The Barenco
three-parameter family of universal two-bit gates provides an alternative and
equally relevant instance of universal quantum gate \cite{B95}. This gate has
a $4\times4$ unitary matrix representation denoted here as $\mathcal{A}%
_{\text{\textrm{universal}}}^{\left(  \text{\textrm{Barenco}}\right)  }\left(
\phi\text{, }\alpha\text{, }\theta\right)  $. With respect to the network's
computational basis $\mathcal{B}_{\mathcal{H}_{2}^{2}}\overset{\text{def}}%
{=}\left\{  \left\vert 00\right\rangle \text{,}\left\vert 10\right\rangle
\text{, }\left\vert 01\right\rangle \text{, }\left\vert 11\right\rangle
\right\}  $, $\mathcal{A}_{\text{\textrm{universal}}}^{\left(
\text{\textrm{Barenco}}\right)  }\left(  \phi\text{, }\alpha\text{, }%
\theta\right)  $ is defined as%
\begin{equation}
\mathcal{A}_{\text{\textrm{universal}}}^{\left(  \text{\textrm{Barenco}%
}\right)  }\left(  \phi\text{, }\alpha\text{, }\theta\right)  \overset
{\text{def}}{=}\left(
\begin{array}
[c]{cc}%
I_{2\times2} & O_{2\times2}\\
O_{2\times2} & A_{2\times2}\left(  \phi\text{, }\alpha\text{, }\theta\right)
\end{array}
\right)  \text{,} \label{bari}%
\end{equation}
with $I_{l\times l}$ denoting the $l\times l$ identity matrix, $O_{m\times k}$
being the $m\times k$ null matrix, and the matrix $A_{2\times2}\left(
\phi\text{, }\alpha\text{, }\theta\right)  $ being defined as
\begin{equation}
A_{2\times2}\left(  \phi\text{, }\alpha\text{, }\theta\right)  \overset
{\text{def}}{=}\left(
\begin{array}
[c]{cc}%
e^{i_{%
\mathbb{C}
}\alpha}\cos(\theta) & -i_{%
\mathbb{C}
}e^{i_{%
\mathbb{C}
}\left(  \alpha-\phi\right)  }\sin(\theta)\\
-i_{%
\mathbb{C}
}e^{i_{%
\mathbb{C}
}\left(  \alpha+\phi\right)  }\sin(\theta) & e^{i_{%
\mathbb{C}
}\alpha}\cos(\theta)
\end{array}
\right)  \text{.}%
\end{equation}
From Eq. (\ref{bari}), we observe that the Barenco gate is characterized by
three parameters $\phi$, $\alpha$, and $\theta$. These parameters, in turn,
are fixed irrational multiples of $\pi$ and of each other. In general, it
happens that \emph{almost }all\emph{\ }two-bit (or, $k$-bits with $k>2$)
quantum gates are universal \cite{D95, L96}. Recall that if an arbitrary
unitary quantum operation can be accomplished with arbitrarily small error
probability by making use of a quantum circuit that only employs gates from
$\mathcal{S}$, then a set of quantum gates $\mathcal{S}$ is known to be
universal. In quantum computing, a relevant set of logic gates is provided by%
\begin{equation}
\mathcal{S}_{\text{\textrm{Clifford}}}\overset{\text{def}}{=}\left\{  \hat
{H}\text{, }\hat{P}\text{, }\hat{U}_{\text{CNOT}}\right\}  \text{.}%
\end{equation}
The set $\mathcal{S}_{\text{\textrm{Clifford}}}$ contains the Hadamard-$\hat
{H}$, the phase-$\hat{P}$ and the CNOT-$U_{\text{CNOT}}$ gates and produces
the so-called Clifford group. As pointed out in Ref. \cite{C98}, this group is
the normalizer $\mathcal{N}\left(  \mathcal{G}_{n}\right)  $ of the Pauli
group $\mathcal{G}_{n}$ in $\mathcal{U}\left(  n\right)  $. While the set of
gates in $\mathcal{S}_{\text{\textrm{Clifford}}}$ suffices to accomplish
fault-tolerant quantum computing, it is not sufficient\textbf{ }to carry out
universal quantum computation. Fortunately, if the gates in $\mathcal{S}%
_{\text{\textrm{Clifford}}}$ are supplemented with the Toffoli gate
\cite{S96}, universal quantum computation can be realized by%
\begin{equation}
\mathcal{S}_{\text{\textrm{universal}}}^{\left(  \text{\textrm{Shor}}\right)
}\overset{\text{def}}{=}\left\{  \hat{H}\text{, }\hat{P}\text{, }\hat
{U}_{\text{CNOT}}\text{, }\hat{U}_{\text{\textrm{Toffoli}}}\right\}  \text{.}
\label{peter}%
\end{equation}
As demonstrated by Shor \cite{S96}, the addition of the Toffoli gate to the
generators of $\mathcal{S}_{\text{\textrm{Clifford}}}$ gives rise to the
universal set of quantum gates $\mathcal{S}_{\text{\textrm{universal}}%
}^{\left(  \text{\textrm{Shor}}\right)  }$ in Eq. (\ref{peter}). An
alternative example of a set of universal logic gates was proposed by Boykin
and collaborators in Refs. \cite{B99, B00}. This different set of gates is
defined as,%
\begin{equation}
\mathcal{S}_{\text{\textrm{universal}}}^{\left(  \text{\textrm{Boykin et al.}%
}\right)  }\overset{\text{def}}{=}\left\{  \hat{H}\text{, }\hat{P}\text{,
}\hat{T}\text{, }\hat{U}_{\text{CNOT}}\text{ }\right\}  \text{.} \label{boy}%
\end{equation}
From a physical realization standpoint, the set of gates in Eq. (\ref{boy}) is
presumably easier to implement experimentally than the set of gates in Eq.
(\ref{peter}) given that the Toffoli gate $\hat{U}_{\text{\textrm{Toffoli}}}$
is a three-qubit gate while the $\pi/8$-quantum gate $\hat{T}$ is a one-qubit gate.

We are now ready to revisit, from a GA language standpoint, the proof of
universality of the set of quantum gates in Eq. (\ref{boy}) as originally
proposed by Boykin and collaborators in Refs. \cite{B99, B00}.

\subsection{GA description of the universality proof}

The universality proof, as originally proposed by Boykin and collaborators in
Refs. \cite{B99, B00}, is quite elegant and relies on two main ingredients.
First, it depends on the local isomorphism between the Lie groups
$\mathrm{SO}(3)$ and $\mathrm{SU}(2)$. Second, it exploits the geometry of
real rotations in three dimensions. In the following, using the rotor group
$\mathrm{Spin}^{+}\left(  3\text{, }0\right)  $ along with the algebra of
bivectors (i.e., $\left[  B_{l}\text{, }B_{m}\right]  =-2\varepsilon
_{lmk}B_{k}$), we shall reconsider the proof in the GA language.

One can use two steps to present the universality proof of $\mathcal{S}%
_{\text{\textrm{universal}}}^{\left(  \text{\textrm{Boykin et al.}}\right)  }$
in Eq. (\ref{boy}). In the first step, one needs to demonstrate that the
Hadamard gate $\hat{H}$ and the $\pi/8$-phase gate $\hat{T}=$ $\hat{\Sigma
}_{3}^{1/4}$ give rise to a \emph{dense} set in the group $\mathrm{SU}(2)$
where, in the GA\ language, we have
\begin{equation}
\hat{\Sigma}_{3}^{\alpha}\overset{\text{def}}{=}\left(
\begin{array}
[c]{cc}%
1 & 0\\
0 & e^{i_{%
\mathbb{C}
}\pi\alpha}%
\end{array}
\right)  \text{, }\hat{\Sigma}_{3}^{\alpha}\left\vert \psi\right\rangle
\leftrightarrow\psi_{\hat{\Sigma}_{3}^{\alpha}}^{\left(  \text{\textrm{GA}%
}\right)  }=\sigma_{3}^{\alpha}\psi\sigma_{3}\text{. }%
\end{equation}
The density of the set $\left\{  \hat{H}\text{, }\hat{T}\right\}  $ implies
that a finite product of $\hat{H}$ and $\hat{T}$ can approximate any element
$\hat{U}_{\mathrm{SU}(2)}\in\mathrm{SU}(2)$ to an suitably chosen degree of
precision. Put differently, it suffices to possess an approximate
implementation of the element $\hat{U}$ with some particular level of
accuracy, when a circuit of quantum gates is employed to realize a suitably
selected unitary operation $\hat{U}$. Assume we use a unitary transformation
$\hat{U}^{\prime}$ to approximate a unitary operation $\hat{U}$. Then, the
so-called approximation error $\varepsilon\left(  \hat{U}\text{, }\hat
{U}^{\prime}\right)  $ is a good measure of the quality of the approximation
of a unitary transformation $\hat{U}$ in terms of $\hat{U}$ \cite{afamm},%
\begin{equation}
\varepsilon\left(  \hat{U}\text{, }\hat{U}^{\prime}\right)  \overset
{\text{def}}{=}\max_{\left\vert \psi\right\rangle }\left\Vert \left(  \hat
{U}-\hat{U}^{\prime}\right)  \left\vert \psi\right\rangle \right\Vert \text{,}%
\end{equation}
with $\left\Vert \psi\right\Vert =\sqrt{\left\langle \psi|\psi\right\rangle }$
denoting the Euclidean norm of $\left\vert \psi\right\rangle $ and
$\left\langle \cdot|\cdot\right\rangle $ being the usual inner product settled
on the complex Hilbert space being considered. In the second step of the
universality proof, it is required to stress that all that is needed for
universal quantum computing is $\hat{U}_{\text{CNOT}}$ and $\mathrm{SU}(2)$
\cite{barenco}. The local isomorphism between $\mathrm{SO}(3)$ and
$\mathrm{SU}(2)$ must be exploited to demonstrate that $\hat{H}$ and $\hat{T}$
give rise to a \emph{dense} set in $\mathrm{SU}(2)$. As a matter of fact,
using the set of gates $\left\{  \hat{H}\text{, }\hat{T}\right\}  $, we can
generate gates that coincide with rotations in $\mathrm{SO}(3$, $%
\mathbb{R}
)$ about two orthogonal axes by angles expressed as irrational multiples of
$\pi$. Examine the following two rotations in $\mathrm{SO}(3)$ specified by
means of rotors in $\mathrm{Spin}^{+}\left(  3\text{, }0\right)  $,%
\begin{equation}
\mathrm{SO}(3)\ni\hat{U}_{\mathrm{SO}(3)}^{\left(  1\right)  }\overset
{\text{def}}{=}e^{i_{%
\mathbb{C}
}\lambda_{1}\pi\hat{n}_{1}\cdot\vec{\Sigma}}\leftrightarrow e^{in_{1}%
\lambda_{1}\pi}\in\mathrm{Spin}^{+}\left(  3\text{, }0\right)  \text{, }%
\hat{U}_{\mathrm{SO}(3)}^{\left(  2\right)  }\overset{\text{def}}{=}e^{i_{%
\mathbb{C}
}\lambda_{2}\pi\hat{n}_{2}\cdot\vec{\Sigma}}\leftrightarrow e^{in_{2}%
\lambda_{2}\pi}\text{, } \label{111}%
\end{equation}
with $\lambda_{1}$, $\lambda_{2}$ being irrational numbers in $%
\mathbb{R}
/%
\mathbb{Q}
$. Let us verify that the two rotations in Eq. (\ref{111}) can be described by
means of a convenient combination of elements belonging to $\left\{  \hat
{H}\text{, }\hat{T}\right\}  $ with $\hat{T}=\hat{\Sigma}_{3}^{1/4}$. Given
that $\mathrm{SU}(2)/%
\mathbb{Z}
_{2}\cong\mathrm{SO}(3)$, it happens to be that
\begin{equation}
\mathrm{Spin}^{+}\left(  3\text{, }0\right)  \ni e^{in_{1}\lambda_{1}\pi
}\leftrightarrow\hat{U}_{\mathrm{SU}(2)}^{\left(  1\right)  }\overset
{\text{def}}{=}\hat{\Sigma}_{3}^{-1/4}\hat{\Sigma}_{1}^{1/4}\in\mathrm{SU}%
(2)\text{ and, }e^{in_{2}\lambda_{2}\pi}\leftrightarrow\hat{U}_{\mathrm{SU}%
(2)}^{\left(  2\right)  }\overset{\text{def}}{=}\hat{H}^{-1/2}\hat{\Sigma}%
_{3}^{-1/4}\hat{\Sigma}_{1}^{\frac{1}{4}}\hat{H}^{1/2}\text{,} \label{m}%
\end{equation}
with $\hat{\Sigma}_{1}^{1/4}=\hat{H}\hat{\Sigma}_{3}^{1/4}\hat{H}$. Exploiting
our findings described in Section III and, in addition, hammering out the
technical details in Refs. \cite{B99, B00}, the rotor representations of
$\hat{U}_{\mathrm{SU}(2)}^{\left(  1\right)  }$ and $\hat{U}_{\mathrm{SU}%
(2)}^{\left(  2\right)  }$ become%
\begin{equation}
\hat{U}_{\mathrm{SU}(2)}^{\left(  1\right)  }\leftrightarrow R_{1}=\frac{1}%
{2}\left(  1+\frac{1}{\sqrt{2}}\right)  \mathbf{-}\frac{1}{2\sqrt{2}}%
i\sigma_{1}+\frac{1}{2}\left(  1-\frac{1}{\sqrt{2}}\right)  i\sigma_{2}%
+\frac{1}{2\sqrt{2}}i\sigma_{3}\text{,} \label{7a}%
\end{equation}
and,%
\begin{equation}
\hat{U}_{\mathrm{SU}(2)}^{\left(  2\right)  }\leftrightarrow R_{2}=\frac{1}%
{2}\left(  1+\frac{1}{\sqrt{2}}\right)  \mathbf{-}\frac{1}{2}\left(  \frac
{1}{2}-\frac{1}{\sqrt{2}}\right)  i\sigma_{1}+\frac{1}{2}i\sigma_{2}+\frac
{1}{2}\left(  \frac{1}{2}-\frac{1}{\sqrt{2}}\right)  i\sigma_{3}\text{,}
\label{7b}%
\end{equation}
respectively. Note that $R_{1}$ and $R_{2}$ in Eqs. (\ref{7a}) and (\ref{7b}),
respectively, denote rotors that belong to $\mathrm{Spin}^{+}\left(  3\text{,
}0\right)  $. Observe that,%
\begin{equation}
e^{in_{k}\lambda_{k}\pi}=\cos\left(  \lambda_{k}\pi\right)  +n_{kx}\sin\left(
\lambda_{k}\pi\right)  i\sigma_{1}+n_{ky}\sin\left(  \lambda_{k}\pi\right)
i\sigma_{2}+n_{kz}\sin\left(  \lambda_{k}\pi\right)  i\sigma_{3}\text{,}%
\end{equation}
for unit vectors $n_{k}$ with $k=1$, $2$. Therefore, putting $e^{in_{1}%
\lambda_{1}\pi}=R_{1}$, we obtain%
\begin{equation}
\cos\left(  \lambda_{1}\pi\right)  =\frac{1}{2}(1+\frac{1}{\sqrt{2}})\text{,
}n_{1y}\sin\left(  \lambda_{1}\pi\right)  =\frac{1}{2}(1-\frac{1}{\sqrt{2}%
})\text{, }n_{1z}\sin\left(  \lambda_{1}\pi\right)  =\frac{1}{2}\frac{1}%
{\sqrt{2}}\text{, }n_{1x}=-n_{1z}\text{.}%
\end{equation}
After some algebraic calculations, the number $\lambda_{1}$ reduces to%
\begin{equation}
\lambda_{1}=\frac{1}{\pi}\cos^{-1}\left[  \frac{1}{2}(1+\frac{1}{\sqrt{2}%
})\right]  \text{.} \label{lambda}%
\end{equation}
Moreover, the unit vector $n_{1}\overset{\text{def}}{=}n_{1x}\sigma_{1}%
+n_{1y}\sigma_{2}+n_{1z}\sigma_{3}$ becomes%
\begin{equation}
n_{1}=\left(  n_{1x}\text{, }n_{1y}\text{, }n_{1z}\right)  =\frac{1}%
{\sqrt{1-\left[  \frac{1}{2}(1+\frac{1}{\sqrt{2}})\right]  ^{2}}}\left(
-\frac{1}{2\sqrt{2}}\text{, }\frac{1}{2}(1-\frac{1}{\sqrt{2}})\text{, }%
\frac{1}{2\sqrt{2}}\right)  \text{ .} \label{BB}%
\end{equation}
Analogously, putting $e^{in_{1}\lambda_{2}\pi}=R_{2}$, we get%
\begin{equation}
\cos\left(  \lambda_{2}\pi\right)  =\frac{1}{2}(1+\frac{1}{\sqrt{2}})\text{,
}n_{2y}\sin\left(  \lambda_{2}\pi\right)  =\frac{1}{2}\text{, }n_{2z}%
\sin\left(  \lambda_{2}\pi\right)  =\frac{1}{2}(\frac{1}{2}-\frac{1}{\sqrt{2}%
})\text{, }n_{2x}=-n_{2z}\text{.}%
\end{equation}
After some additional algebraic calculations, we have that $\lambda
_{2}=\lambda_{1}$. Therefore, from Eq. (\ref{lambda}), $\lambda_{2}$ becomes
\begin{equation}
\lambda_{2}=\frac{1}{\pi}\cos^{-1}\left[  \frac{1}{2}(1+\frac{1}{\sqrt{2}%
})\right]  \text{.}%
\end{equation}
The unit vector $n_{2}\overset{\text{def}}{=}n_{2x}\sigma_{1}+n_{2y}\sigma
_{2}+n_{2z}\sigma_{3}$, instead, reduces to%
\begin{equation}
n_{2}=\left(  n_{2x}\text{, }n_{2y}\text{, }n_{2z}\right)  =\frac{1}%
{\sqrt{1-\left[  \frac{1}{2}(1+\frac{1}{\sqrt{2}})\right]  ^{2}}}\left(
-\frac{1}{2}(\frac{1}{2}-\frac{1}{\sqrt{2}})\text{, }\frac{1}{2}\text{, }%
\frac{1}{2}(\frac{1}{2}-\frac{1}{\sqrt{2}})\text{ }\right)  \text{.}
\label{BBB}%
\end{equation}
As a consistency check, we can verify that Eqs. (\ref{BB}) and (\ref{BBB})
imply that $n_{1}\cdot n_{2}=0$. Since $\lambda_{1}=\lambda_{2}=\lambda\in%
\mathbb{R}
/%
\mathbb{Q}
$, any phase factor $e^{i_{%
\mathbb{C}
}\phi}$ can be approximately described by $e^{i_{%
\mathbb{C}
}n\lambda\pi}$ for some $n\in%
\mathbb{N}
$ ,%
\begin{equation}
e^{i_{%
\mathbb{C}
}\phi}\approx e^{i_{%
\mathbb{C}
}n\lambda\pi}\text{.} \label{mmm}%
\end{equation}
Eqs. (\ref{m}) and (\ref{mmm}) imply that we possess at least two dense
subsets of $\mathrm{SU}(2$, $%
\mathbb{C}
)$. They are characterized by $e^{in_{1}\alpha}$ and $e^{i\beta n_{2}}$ where,%
\begin{equation}
\alpha\approx\lambda\pi l\text{ (\textrm{mod}}2\pi\text{) and, }\beta
\approx\lambda\pi l\text{ (\textrm{mod}}2\pi\text{),}%
\end{equation}
with $l\in%
\mathbb{N}
$. We observe that we are allowed to express any element $\hat{U}%
_{\mathrm{SU}(2)}$ $\in$ $\mathrm{SU}(2$, $%
\mathbb{C}
)$ as,%
\begin{equation}
\hat{U}_{\mathrm{SU}(2)}=e^{i_{%
\mathbb{C}
}\phi\hat{n}\cdot\vec{\Sigma}}\leftrightarrow e^{in\phi}=e^{in_{1}\alpha
}e^{in_{2}\beta}e^{in_{1}\gamma}\text{,} \label{jj}%
\end{equation}
given that $n_{1}$ and $n_{2}$ in Eqs. (\ref{BB}) and (\ref{BBB}),
respectively, are orthogonal (unit) vectors. Interestingly, note that the
decomposition in Eq. (\ref{jj}) can be regarded as the analogue of the Euler
rotations about three orthogonal vectors. Expanding the left-hand-side and the
right-hand-side of the second relation in Eq. (\ref{jj}), we get%
\begin{equation}
e^{in\phi}=\cos(\phi)+in\sin(\phi)\text{.} \label{cx}%
\end{equation}
and,%
\begin{equation}
e^{in_{1}\alpha}e^{in_{2}\beta}e^{in_{1}\gamma}=\left[  \cos(\alpha
)+in_{1}\sin(\alpha)\right]  \left[  \cos(\beta)+in_{2}\sin(\beta)\right]
\left[  \cos(\gamma)+in_{1}\sin(\gamma)\right]  \text{,} \label{yes}%
\end{equation}
respectively. Recollecting that $n_{1}n_{2}=n_{1}\cdot n_{2}+n_{1}\wedge
n_{2}$ and, in addition, that the unit vectors $n_{1}$ and $n_{2}$ are
orthogonal, we obtain%
\begin{equation}
n_{1}n_{2}=-n_{2}n_{1}\text{.} \label{too}%
\end{equation}
Furthermore, keeping in mind the following trigonometric relations%
\begin{equation}
\sin\left(  \alpha\pm\beta\right)  =\sin(\alpha)\cos(\beta)\pm\cos(\alpha
)\sin(\beta)\text{ and, }\cos\left(  \alpha\pm\beta\right)  =\cos(\alpha
)\cos(\beta)\mp\sin(\alpha)\sin(\beta)\text{, } \label{xxx}%
\end{equation}
additional manipulation of Eq. (\ref{yes}) along with the employment of Eqs.
(\ref{too})\ and (\ref{xxx}), yields%
\begin{equation}
e^{in\phi}=\cos(\beta)\cos\left(  \alpha+\gamma\right)  +\cos(\beta
)\sin\left(  \alpha+\gamma\right)  in_{1}+\sin(\beta)\cos\left(  \gamma
-\alpha\right)  in_{2}+\sin(\beta)\sin\left(  \gamma-\alpha\right)
n_{1}\wedge n_{2}\text{.} \label{cxx}%
\end{equation}
Putting Eq. (\ref{cx}) equal to Eq. (\ref{cxx}), we finally get%
\begin{equation}
\cos(\phi)=\cos(\beta)\cos\left(  \alpha+\gamma\right)  \label{A}%
\end{equation}
and,%
\begin{equation}
n\sin(\phi)=\cos(\beta)\sin\left(  \alpha+\gamma\right)  n_{1}+\sin(\beta
)\cos\left(  \gamma-\alpha\right)  n_{2}-i\sin(\beta)\sin\left(  \gamma
-\alpha\right)  \left(  n_{1}\wedge n_{2}\right)  \text{.} \label{B}%
\end{equation}
In closing, the parameters $\alpha$, $\beta$ and $\gamma$ can be obtained once
Eqs. (\ref{A}) and (\ref{B}) are inverted for any element in $\mathrm{SU}(2)$.
Then, exploiting the fact that $\hat{U}_{\text{CNOT}}$ and $\mathrm{SU}(2)$
give rise to a universal basis of quantum gates for quantum computation
\cite{barenco}, the GA\ version of the universality proof as originally
proposed in Refs. \cite{B99, B00} is achieved.

As evident from our work, we reiterate that the GA language offers a very neat
and solid technique for encoding rotations which is significantly more
powerful than computing with matrices. Moreover, as apparent from several
applications of GA\ in mathematical physics, a conceptually relevant feature
of GA appears. Namely, vectors (i.e., grade-$1$ multivectors), planes (i.e.,
grade-$2$ multivectors), and the operators acting on them (i.e., rotors $R$
and bivectors $B$ in our case) are elements that belong to the very same
geometric Clifford algebra.

\section{Concluding remarks}

In this paper, we revisited the usefulness of GA techniques in two particular
applications in QIS. In our first application, we offered an instructive
MSTA\ characterization of one- and two-qubit quantum states together with a
MSTA description of one- and two-qubit quantum computational gates. In our
second application, instead, we used the findings of our first application
together with the GA characterization of the Lie algebras $\mathrm{SO}\left(
3\right)  $ and\textrm{ }$\mathrm{SU}\left(  2\right)  $ in terms of the rotor
group $\mathrm{Spin}^{+}\left(  3\text{, }0\right)  $ formalism to revisit the
proof of universality of quantum gates as originally proposed by Boykin and
collaborators in Refs. \cite{B99,B00}. Our main conclusions are two. First of
all, in agreement with what was stressed in Ref. \cite{laser}, we point out
that the MSTA approach gives rise to a useful conceptual unification in which
multivectors in real space provide a unifying setting for both the complex
qubit space and the complex space of unitary operators acting on them. Second
of all, the GA perspective on rotations in terms of the rotor group
$\mathrm{Spin}^{+}\left(  3\text{, }0\right)  $ undoubtedly introduces both
computational and conceptual benefits compared to ordinary vector and matrix
algebra approaches.

In the following, we present some additional remarks related to our proposed
use of GA\ in QIS.

\begin{enumerate}
\item[{[i]}] In the ordinary formulation of quantum computing, the essential
operation is represented by the tensor product\emph{\ }\textquotedblleft%
$\otimes$\textquotedblright. The basic operation in the GA approach to quantum
computation, instead, is the geometric (Clifford) product. Unlike tensor
products, geometric products have transparent geometric interpretations.
Indeed, using the geometric product, one can use a vector $\left(  \sigma
_{1}\right)  $ and a square $\left(  \sigma_{2}\sigma_{3}\right)  $ to form a
cube $\left(  \sigma_{1}\sigma_{2}\sigma_{3}\right)  $. Alternatively, from
two vectors $\left(  \sigma_{1}\text{ and }\sigma_{2}\right)  $, one can
generate an oriented square $\left(  \sigma_{1}\sigma_{2}\right)  $. Also,
among many more possibilities, one can form a square $\left(  \sigma_{2}%
\sigma_{3}\right)  $ from a cube $\left(  \sigma_{1}\sigma_{2}\sigma
_{3}\right)  $ and a vector $\left(  \sigma_{1}\right)  $.

\item[{[ii]}] (Complex) entangled quantum states in ordinary formulations of
quantum computing are replaced by (real) multivectors with a clear geometric
interpretation within the GA language. For instance, a general multivector $M$
in $\mathfrak{Cl}(3)$ is a linear combination of blades, geometric products of
different basis vectors supplemented by the identity $\mathbf{1}$ (basic
oriented scalar),%
\begin{equation}
M\overset{\text{def}}{=}M_{0}\mathbf{1}+\sum_{j=1}^{3}M_{j}\sigma_{j}%
+\sum_{j<k}M_{jk}\sigma_{j}\sigma_{k}+M_{123}\sigma_{1}\sigma_{2}\sigma
_{3}\text{, }%
\end{equation}
with $j$, $k=1$, $2$, $3$. In this context, entangled quantum states are
viewed as GA multivectors that are nothing but bags of shapes (i.e., points,
$1$; lines, $\sigma_{j}$; squares, $\sigma_{j}\sigma_{k}$; cubes, $\sigma
_{1}\sigma_{2}\sigma_{3}$).

\item[{[iii]}] One of the key aspect of GA, that we emphasized in Ref.
\cite{mancio11} and reiterated in the above point [ii] of this paper, is that
\ (complex) operators and operands (i.e., states) are elements of the same
(real) space in the GA setting. This fact, in turn, is at the root of the
increasing number of works advocating for the use of online calculators
capable of performing quantum computing operations based on geometric algebra
\cite{hild22,alves22,J23,J24}. We are proud to see that our original work in
Ref. \cite{mancio11} has had an impact on these more recent works supporting
the use of GA-based online calculators in QIS.

\item[{[iv]}] Describing and, to a certain extent, understanding the
complexity of quantum motion of systems in entangled quantum states remains a
truly fascinating problem in quantum physics with several open issues. In QIS,
the notion of quantum gate complexity, defined for quantum unitary operators
and regarded as a measure of the computational work necessary to accomplish a
given task, is a significant complexity measure \cite{NIELSEN}. It would be
intriguing to explore if the conceptual unification between (complex) spaces
of \ quantum states and of quantum unitary operators acting on such states
offered by MSTAs can allow for the possibility of yielding a unifying
mathematical setting in which complexities of both quantum states and quantum
gates are defined for quantities that belong to the same real multivectorial
space. Note that geometric reasoning demands the reality of the multivectorial
space. Moreover, we speculate that this conceptual unification may happen to
be very beneficial with respect to the captivating link between quantum gate
complexity and complexity of the motion on a suitable Riemannian manifold of
multi-qubit unitary transformations given by Nielsen and collaborators in
Refs. \cite{nc, nc1}.
\end{enumerate}

\smallskip

In view of our quantitative findings revisited here, along with our more
speculative considerations, we have reason to believe that the use of
geometric Clifford algebras in QIS together with its employment in the
characterization of quantum gate complexity appears to be deserving of further
theoretical explorations
\cite{brown19,chapman18,ruan21,cafaroprd22,cafaropre22}. Moreover, motivated
by our revisitation of GA\ methods in quantum information science together
with our findings appeared in Refs. \cite{pra22,cqg23}, we think that the
application of the GA language (with special emphasis on the concept of
rotation) can be naturally extended (for gaining deeper physical insights) to
the analysis of the propagation of light with maximal degree of coherence
\cite{pra22,wolf07,loudon00} and, in addition, to the characterization of the
geometry of quantum evolutions
\cite{cqg23,uzdin12,campaioli19,dou23,huang24,musz13,mandarino18}.

We have limited our discussion in this paper to the universality for qubit
systems. However, it would be fascinating to explore the usefulness of the GA
language in the context of universality problems for higher-dimensional
systems, i.e. qudits \cite{brylinski02,saw17,saw17B,saw22}. Interestingly, for
quantum computation by means of qudits \cite{cafaroqudit}, a universal set of
gates is specified by all one-qudit gates together with any additional
entangling two-qudit gate \cite{brylinski02} (that is, a gate that does not
map separable states onto separable states). Finally, it would be of
theoretical interest exploiting our work as a preliminary starting point from
which extending the use of GA techniques from Grover's algorithm with qubits
\cite{alves10,chap12,caf17} to Grover's algorithm with qudits \cite{niko23}.
For the time being, we leave these intriguing scientific explorations as
future works.

\smallskip

\begin{acknowledgments}
C.C. thanks Professor Ariel Caticha for having introduced geometric algebra to
him during his PhD in Physics at the University at Albany-SUNY (2004-2008,
USA). C.C. is also grateful to Professor Stefano Mancini for having welcomed
the use of geometric algebra techniques in quantum computing during his
postdoctoral experience at the Physics-Division of the University of Camerino
(2009-2011, Italy). The authors thank four anonymous reviewers for useful
comments leading to an improved version of this manuscript. Any opinions,
findings and conclusions or recommendations expressed in this material are
those of the author(s) and do not necessarily reflect the views of their home Institutions.
\end{acknowledgments}

\bigskip\pagebreak

\appendix

\section{From the algebra of physical space to spacetime algebra}

In this Appendix, we present essential elements of the algebra of physical
space $\mathfrak{cl}(3)$ together with the spacetime Clifford algebra
$\mathfrak{cl}(1,3)$.

\subsection{Algebra of physical space $\mathfrak{cl}(3)$}

Geometric algebra is Clifford's generalization of complex numbers and
quaternion algebra to vectors in arbitrary dimensions. The result is a
formalism in which elements of any grade (including scalars, vectors, and
bivectors) can be added or multiplied together is called \emph{geometric
algebra}. For two vectors $a$ and $b$ the \emph{geometric product} is the sum
of an inner product and an outer product given by
\begin{equation}
\vec{a}\text{ }\vec{b}\overset{\text{def}}{=}\vec{a}\cdot\vec{b}+\vec{a}%
\wedge\vec{b}.
\end{equation}
Geometric product is associative and has the crucial feature of being invertible.

In three dimensions where $\vec{a}$ and $\vec{b}$ are three-dimensional
vectors the inner product is a scalar (grade-$0$ multivector) and the outer
product is a bivector (grade-$2$ multivector). Considering a right-handed
frame of orthonormal basis vectors ${\vec{e}_{1},\vec{e}_{2},\vec{e}_{3}}$, we
have
\begin{equation}
\vec{e}_{l}\vec{e}_{m}=\vec{e}_{l}\cdot\vec{e}_{m}+\vec{e}_{l}\wedge\vec
{e}_{m}=\delta_{lm}+\varepsilon_{lmk}i\vec{e}_{k} \label{q1}%
\end{equation}
where $i\overset{\text{def}}{=}\vec{e}_{1}\vec{e}_{2}\vec{e}_{3}$ is the
pseudoscalar which is a trivector (grade-$3$ multivector) and it is the
directed unit volume element. Pauli spin matrices also satisfy a relation
similar to Eq. (\ref{q1}). Thus, Pauli spin matrices form a matrix
representation of the geometric algebra of physical space. The geometric
algebra of three-dimensional physical space (APS) is spanned by one scalar,
three vectors, three bivectors, and one trivector which defines a graded
linear space of $8=2^{3}$ dimensions called $\mathfrak{cl}(3)$,
\begin{equation}
\mathfrak{cl}(3)\overset{\text{def}}{=}\mathrm{Span}\left\{  {1}\text{{; }%
}{\vec{e}_{1},\vec{e}_{2},\vec{e}_{3}}\text{{; }}{\vec{e}_{1}\vec{e}_{2}%
,\vec{e}_{2}\vec{e}_{3},\vec{e}_{3}\vec{e}_{1}}\text{{; }}{\vec{e}_{1}\vec
{e}_{2}\vec{e}_{3}}\right\}  \text{.}%
\end{equation}
We also have that $i^{2}=-1$. Since $\vec{e}_{1},\vec{e}_{2},\vec{e}_{3}$ in
the Pauli algebra are given by Pauli spin matrices, we can write $i^{\dagger
}=-i$ where \textquotedblleft$\dagger$\textquotedblright\ is the Hermitian
conjugate in the Pauli spin matrices. This can give a geometric interpretation
of $i$ in quantum mechanics. For a general multivector $\bar{M}$ in
$\mathfrak{cl}(3)$ we can write
\begin{equation}
\bar{M}=\alpha+\vec{a}+B+\beta i
\end{equation}
where $\alpha$ and $\beta$ are scalars denoted by ${\left\langle \bar
{M}\right\rangle }_{0}$, $\vec{a}$ is a vector denoted by ${\left\langle
\bar{M}\right\rangle }_{1}$, $B$ is a bivector denoted by ${\left\langle
\bar{M}\right\rangle }_{2}$, and $i$ is a trivector denoted by ${\left\langle
\bar{M}\right\rangle }_{3}$
\begin{equation}
\bar{M}=\underset{k=0,1,2,3}{\sum}\left\langle \bar{M}\right\rangle
_{k}=\left\langle \bar{M}\right\rangle _{0}+\left\langle \bar{M}\right\rangle
_{1}+\left\langle \bar{M}\right\rangle _{2}+\left\langle \bar{M}\right\rangle
_{3} \label{n4}%
\end{equation}
A bivector $B$ can be written as $B=i\vec{b}$, where $\vec{b}$ is a vector.
Substituting this into Eq. (\ref{n4}) we get
\begin{equation}
\bar{M}=\alpha+\vec{a}+i\vec{b}+i\beta=\text{scalar+ vector+ bivector+
trivector.} \label{n6}%
\end{equation}
Eq. (\ref{n6} ) can be rearranged as $($complex scalar + complex
vector$)=(\alpha+i\beta)+(\vec{a}+i\vec{b})$ which can be written as
\begin{equation}
\bar{M}=\left\langle \bar{M}\right\rangle _{cs}+\left\langle \bar
{M}\right\rangle _{cv}=\left[  \left\langle \bar{M}\right\rangle
_{rs}+\left\langle \bar{M}\right\rangle _{is}\right]  +\left[  \left\langle
\bar{M}\right\rangle _{rv}+\left\langle \bar{M}\right\rangle _{iv}\right]
=M^{0}+\vec{M}%
\end{equation}
where $\left\langle \bar{M}\right\rangle _{cs}$ is the sum of real and
imaginary scalar components,
\begin{equation}
\left\langle \bar{M}\right\rangle _{cs}\overset{\text{def}}{=}M^{0}%
=\left\langle \bar{M}\right\rangle _{rs}+\left\langle \bar{M}\right\rangle
_{is}%
\end{equation}
while $\left\langle \bar{M}\right\rangle _{cv}$ consists of real and imaginary
vector components
\begin{equation}
\left\langle \bar{M}\right\rangle _{cv}\overset{\text{def}}{=}M^{0}%
=\left\langle \bar{M}\right\rangle _{rv}+\left\langle \bar{M}\right\rangle
_{iv}.
\end{equation}
This is referred to as a paravector and it is used by Baylis to model
spacetime. More details can be found in Refs.
\cite{baylis04,baylis89,vaz18,vaz19}.

Two involutions can be used, the \emph{reversion} or \emph{Hermitian adjoint}
\textquotedblleft$\dag$\textquotedblright\ and the \emph{spatial reverse} or
\emph{Clifford conjugate} \textquotedblleft$\ddag$\textquotedblright. For an
arbitrary element multivector $\bar{M}=\alpha+\vec{a}+i\vec{b}+i\beta$, these
involutions are defined as,
\begin{equation}
\bar{M}^{\dag}\overset{\text{def}}{=}\alpha+\vec{a}-i\vec{b}-i\beta\text{ and,
}\bar{M}^{\ddag}\overset{\text{def}}{=}\alpha-\vec{a}-i\vec{b}+i\beta\text{.}%
\end{equation}
We use here the following notation $\underline{M}\overset{\text{def}}{=}%
\bar{M}^{\ddag}$. Useful identities are,%
\begin{equation}
\left\langle \underline{M}\right\rangle _{rs}=\frac{1}{4}\left[  \underline
{M}+\underline{M}^{\dag}+\underline{M}^{\ddag}+\left(  \underline{M}^{\dag
}\right)  ^{\ddag}\right]  \text{, }\left\langle \underline{M}\right\rangle
_{rv}=\frac{1}{4}\left[  \underline{M}^{\ddag}+\left(  \underline{M}^{\dag
}\right)  ^{\ddag}-\underline{M}-\underline{M}^{\dag}\right]  \text{,}%
\end{equation}%
\begin{equation}
\left\langle \underline{M}\right\rangle _{is}=\frac{1}{4}\left[  \underline
{M}-\underline{M}^{\dag}+\underline{M}^{\ddag}-\left(  \underline{M}^{\dag
}\right)  ^{\ddag}\right]  \text{, }\left\langle \underline{M}\right\rangle
_{iv}=\frac{1}{4}\left[  \underline{M}^{\dag}-\underline{M}+\underline
{M}^{\ddag}-\left(  \underline{M}^{\dag}\right)  ^{\ddag}\right]  \text{.}
\label{w13}%
\end{equation}
Moreover, an important algebra of physical space vector is the vector
derivatives $\bar{\partial}$ and $\underline{\partial}\overset{\text{def}}{=}$
$\bar{\partial}^{\ddag}$ defined by,
\begin{equation}
\bar{\partial}=\bar{e}_{\mu}\partial^{\mu}=c^{-1}\partial_{t}-\vec{\nabla
}\text{ and, }\underline{\partial}=\text{\b{e}}^{\mu}\partial_{\mu}%
=c^{-1}\partial_{t}+\vec{\nabla}\text{.}%
\end{equation}
Finally, the d'Alambertian differential wave scalar operator $\square
_{\mathfrak{cl}(3)}$ in the APS\ formalism is,
\begin{equation}
\square_{\mathfrak{cl}\mathrm{l}(3)}\overset{\text{def}}{=}\underline
{\partial}\overline{\partial}=\bar{e}_{\mu}^{\nu}\text{\b{e}}^{\nu}%
\partial^{\mu}\partial_{\nu}=\delta_{\nu}^{\mu}\partial^{\mu}\partial_{\nu
}=\partial^{\mu}\partial_{\mu}\equiv\partial^{2}=c^{-2}\partial_{t}^{2}%
-\vec{\nabla}^{2}\text{.}%
\end{equation}
It describes lightlike traveling waves. For additional technical details on
the algebra of physical space $\mathfrak{cl}(3)$, we refer to Ref \cite{dl}.
In the next subsection, we present elements of the spacetime algebra
$\mathfrak{cl}(1,3)$.

\subsection{Spacetime algebra $\mathfrak{cl}(1,3)$}

The spacetime algebra (STA) is constructed based on four orthonormal basis
vectors ${\gamma_{0},\gamma_{1},\gamma_{2},\gamma_{3}}$ where $\gamma_{0}$ is
timelike and $\gamma_{1},\gamma_{2},\gamma_{3}$ are spacelike vectors and form
a right-handed orthonormal basis set such that
\begin{equation}
{\gamma_{0}}^{2}=1,\ \ \gamma_{0}.\gamma_{i}=0,\ \ \gamma_{i}\cdot\gamma
_{j}=-\delta_{ij};\ \ i,j=1,2,3
\end{equation}
which can be summarized as
\begin{equation}
\gamma_{\mu}\cdot\gamma_{\nu}=\eta_{\mu\nu}=\mathrm{diag}(+---);\ \mu
,\nu=0,...,3.
\end{equation}
There are two types of bivectors given by
\begin{equation}
(\gamma_{i}\wedge\gamma_{j})^{2}=-{\gamma_{i}}^{2}{\gamma_{j}}^{2}=-1\text{,
and }(\gamma_{i}\wedge\gamma_{0})^{2}=-{\gamma_{i}}^{2}{\gamma_{0}}%
^{2}=1\text{.}%
\end{equation}
Finally, there is the grade-$4$ pseudoscalar is defined by
\begin{equation}
i\overset{\text{def}}{=}\gamma_{0}\gamma_{1}\gamma_{2}\gamma_{3}\text{.}%
\end{equation}
The spacetime algebra, $\mathfrak{cl}(1,3)$ has $16$ terms which includes one
scalar, four vectors $({\gamma_{\mu}})$, six bivectors $({\gamma_{\mu}%
\wedge\gamma_{\nu}})$, four trivectors $({i\gamma_{\mu}})$, and one
pseudoscalar $(i=\gamma_{0}\gamma_{1}\gamma_{2}\gamma_{3})$ which give
$2^{4}=16$ dimensional STA. Therefore, a basis for the spacetime algebra is
given by
\begin{equation}
\left\{  1,\gamma_{\mu},\gamma_{\mu}\wedge\gamma_{\nu},i\gamma_{\mu
},i\right\}  \text{.}%
\end{equation}
A general element is written as
\begin{equation}
M\overset{\text{def}}{=}{\sum\limits_{k=0}^{4}}\left\langle M\right\rangle
_{k}=\alpha+a+B+ib+i\beta\text{,}%
\end{equation}
where $\alpha$ and $\beta$ are scalars, $a$ and $b$ are vectors and $B$ is a
bivector. The vector generators of spacetime algebra satisfy
\begin{equation}
\gamma_{\mu}\gamma_{\nu}+\gamma_{\nu}\gamma_{\mu}=2\eta_{\mu\nu}.
\end{equation}
These relations indicate that the Dirac matrices are a representation of
spacetime algebra and Minkowski metric tensor's nonzero terms are $(\eta_{00}%
$, $\eta_{11}$, $\eta_{22}$, $\eta_{33})=(1$, $-1$, $-1$, $-1)$. A map between
a general spacetime vector $a=a^{\mu}\gamma_{\mu}$ and the even subalgebra of
the STA $\mathfrak{cl}^{+}(1,3)$ when $\gamma_{0}$ is the future-pointing
timelike unit vector is given by
\begin{equation}
a\gamma_{0}=a_{0}+\vec{a}%
\end{equation}
where
\begin{equation}
a_{0}=a\cdot\gamma_{0}\text{, }\ \vec{a}=a\wedge\gamma_{0}%
\end{equation}
and $\vec{a}$ is an ordinary spatial vector in three dimensions which can be
interpreted as a spacetime bivector. Since the metric is given by $\eta
_{\mu\nu}\overset{\text{def}}{=}\mathrm{diag}(+---)$, the matrix has no zero
eigenvalues and a trace equal to $-2$ which is in agreement with
$\mathfrak{cl}(1,3)$. If instead, the metric is chosen such that the trace is
$+2$, the algebra associated with that would be $\mathfrak{cl}(3,1)$ which is
not isomorphic to $\mathfrak{cl}(1,3)$. In geometric algebra, the pseudoscalar
which is the highest-grade element determines the metric. For the spacetime
Lorentz group, the pseudoscalar satisfies $i^{2}=-1$. Since $n=4$ for this
space, $i$ anticommutes with odd-grade and commutes with even-grade
multivectors of the algebra
\begin{equation}
iP=\pm Pi
\end{equation}
where $(+)$ refers to the case when $P$ is even and $(-)$ is the case when $P$
is odd. An important spacetime vector derivative $\nabla$ is defined by
\begin{equation}
\nabla\overset{\text{def}}{=}\gamma^{\mu}\partial_{\mu}\equiv\gamma^{0}%
c^{-1}\partial_{t}+\gamma^{i}\partial_{i}.
\end{equation}
Post-multiplying by $\gamma^{0}$ gives
\begin{equation}
\nabla\gamma_{0}=c^{-1}\partial_{t}+\gamma^{i}\gamma_{0}\partial_{i}%
=c^{-1}\partial_{t}-\vec{\nabla}%
\end{equation}
where $\vec{\nabla}$ is the usual derivative defined in vector algebra.
Multiplying the spacetime vector derivative by $\gamma^{0}$ gives
\begin{equation}
\gamma_{0}\nabla=c^{-1}\partial_{t}+\vec{\nabla}\text{.}%
\end{equation}
Finally, we notice that the spacetime vector derivative satisfies the
following relation
\begin{equation}
\square=(\gamma_{0}\nabla)(\nabla\gamma_{0})=c^{-2}\partial_{t}^{2}%
-\vec{\nabla}^{2}%
\end{equation}
which is the d'Alembert operator used in the description of lightlike
traveling waves. Additional technical details on the spacetime Clifford
algebra $\mathfrak{cl}(1,3)$ can be found in \cite{dl}.

\end{document}